\documentclass[12pt]{article}
\usepackage{amsmath}
\usepackage{amssymb}
\usepackage{amsthm}
\usepackage{mathrsfs}
\usepackage{amscd}
\usepackage[pdftex]{hyperref}
\usepackage{multirow}
\usepackage{graphicx}
\usepackage{comment}
\usepackage{color}
\usepackage{array}
\usepackage{indentfirst}
\usepackage{enumerate}
\usepackage{natbib}
\usepackage{titlesec}
\usepackage[titletoc]{appendix}
\usepackage{subfigure}

\usepackage{caption}
\usepackage{algorithm}
\usepackage{algorithmicx}
\usepackage{algpseudocode}

\usepackage{subfigure}
\usepackage{dsfont}
\usepackage{booktabs}

\newcommand{\PreserveBackslash}[1]{\let\temp=\\#1\let\\=\temp}

\newcommand{\bas}{\begin{eqnarray*}}
\newcommand{\eas}{\end{eqnarray*}}
\newcommand{\ba}{\begin{eqnarray}}
\newcommand{\ea}{\end{eqnarray}}
\newcommand{\bit}{\begin{itemize}}
\newcommand{\eit}{\end{itemize}}

\newcommand{\bs}[1]{\boldsymbol{ #1 }}
\newcommand{\mb}[1]{\mathbf{ #1 }}

\newcommand*{\mx}{\mb{X}}

\newcommand*{\my}{\mb{Y}}

\newcommand{\mi}{\mb{I}}

\newcommand{\cH}{\mathcal{H}}

\newtheorem{theorem}{Theorem}

\theoremstyle{definition}

\newtheorem{example}{Example}[section]

\usepackage{nameref}

\usepackage{xr}

\makeatletter
\newcommand*{\addFileDependency}[1]{
	\typeout{(#1)}
	\@addtofilelist{#1}
	\IfFileExists{#1}{}{\typeout{No file #1.}}
}
\makeatother

\begin{document}
{
\title{A Partially Functional Linear Modeling Framework for Integrating  Genetic, Imaging, and 
Clinical Data} 
\vspace{-40pt}
\date{}
\maketitle
\begin{center}
  \author{\large Ting Li$^{*1}$, Yang Yu$^{*2}$,  J. S.  Marron$^{2,3}$, and   Hongtu Zhu$^{2,3,4,5,6}$  \\
  $^{1}$School of Statistics and Management, Shanghai University of Finance and Economics, Shanghai, China 
  \\ 
  Departments of  $^{2}$Statistics, $^{3}$Biostatistics, $^{4}$Genetics, and $^{5}$Computer Science and $^{6}$Biomedical Research Imaging Center, University of North Carolina at Chapel Hill, Chapel Hill }
  \footnote{
$^*$These authors contributed equally: Ting Li and Yang Yu.   Address for correspondence: Hongtu Zhu, Ph.D., Email: htzhu@email.unc.edu. 
Data used in preparation of this article were obtained from the Alzheimer's Disease
Neuroimaging Initiative (ADNI) database (adni.loni.usc.edu). As such, the investigators within the ADNI contributed to the design and implementation of ADNI and/or provided data but did not participate in analysis or writing of this report. A complete listing of ADNI investigators can be found at: \url{http://adni.loni.usc.edu/wp-content/uploads/how_to_apply/ADNI_Acknowledgement_List.pdf}.}

\end{center}
\newpage

	\begin{abstract}
	This paper is motivated by the joint analysis of  genetic, imaging, and clinical (GIC) data collected in the Alzheimer's Disease Neuroimaging Initiative (ADNI) study. We propose  a regression framework based on  partially functional linear regression models to  map high-dimensional GIC-related pathways for Alzheimer's Disease (AD).
	We develop a joint model selection and estimation procedure by embedding imaging data in the reproducing kernel Hilbert space and imposing the $\ell_0$ penalty for the coefficients of genetic variables.
	We apply the proposed method to the ADNI dataset to identify important features from tens of thousands of genetic polymorphisms (reduced from millions using a preprocessing step)
	and study the effects of a certain set of informative genetic variants and the baseline hippocampus surface on 
	thirteen future cognitive scores
	measuring different aspects of cognitive function.
	We explore the shared and different heritablity patterns of these cognitive scores.
	Analysis results suggest that both the hippocampal and genetic data have heterogeneous effects on different scores, with the trend that
	the value of both hippocampi is negatively associated with the severity of cognition deficits. Polygenic effects are observed for all the thirteen cognitive scores. 
	The well-known APOE4 genotype only explains a small part of the cognitive function.
	Shared genetic etiology exists,
	however, greater genetic heterogeneity exists within disease classifications after accounting for the baseline diagnosis status.
	These analyses 
	are useful in further investigation of functional mechanisms for AD evolution.
\end{abstract}

\noindent%

\section{Introduction}
\label{sec:PFL-intro}
Alzheimer's disease (AD) is a chronic neurodegenerative disease that causes degeneration of brain cells and decline in thinking, behavioral and social skills. It 
involves cognitive impairment with substantial between-patient variability in clinical presentation as well as the burden and distribution of pathology. 
Such clinicopathologic heterogeneity is both challenges and opportunities for carrying out systematic and biomarker-based studies to refine our understanding of AD biology, diagnosis, and management \citep{duong2022dissociation}.
AD has complex pathophysiological mechanisms which are not completely understood.
The advances in biomarker identification, including genetic  and imaging data, may improve the identification of individuals at risk for AD before symptom onset.

The primary aim  of this study is to  use 
Genetic, Imaging, and Clinical (GIC) variables from 
the ADNI study to map the biological pathways of  
AD related phenotypes of interest (e.g., cognition, intelligence,  disease stage, impairment score, and progression status) \citep{sudlow2015uk,elliott2018genome}. 
It may provide insights into the biological process of  
brain development, healthy aging, and disease progress. 
For instance, it is great interest to
integrate GIC  to  elucidate the environmental, social, and genetic etiologies of intelligence and to delineate  the foundation of intelligence
differences in brain structure and functioning  \citep{deary2021genetic}. 
Moreover, 
many brain-related  disorders including AD are often caused by  
a combination of multiple genetic and environmental factors, while being   the  endpoints  of abnormality  of brain structure and function  \citep{shen2019brain,Zhao2019, knutson2020implicating}.
A thorough understanding of such  neuro-biological pathways may lead to the identification of  possible hundreds of risk genes,  environmental risk factors,  and    brain structure and function that underline  brain disorders.   
Once such identification has been accomplished, it is possible to detect 
these risk genes and factors and brain abnormalities early enough  to make a real difference in outcome and to develop their related treatments, ultimately preventing the onset of brain-related disorders and reducing their severity. 

We extract clinical, imaging and genetic variables from the ADNI study.
It includes cognitive scores for quantifying behavior deficits, ultra-high dimensional genetic covariates, other demongraphic covariates at baseline and brain structures using brain imaging.
As previous studies have shown that the hippocampus is particularly vulnerable to AD pathology
and has become a major focus in AD \citep{braak1998evolution},
we characterize the exposure of interest, hippocampal shape, by the left/right hippocampal morphometry surface data as a 100$\times $150 matrix.
We give a detailed data description in Section \ref{sec:PFL-DataDes}.
Exploring how human brains and genetics connect to human behavior is a central goal in medical studies. We are interested in how hippocampal shape and genetics are associated with future cognition deficits in Alzheimer's study.
The special data structure of these GIC variables presents new challenges
for mapping the GIC pathway. First,  conventional statistical tools that deal with scalar exposure 
are not applicable to 2D high-dimensional hippocampal imaging measures. Second, the dimension of the genetic covariates is much larger than the sample size.
An effective statistical method which can
exploit the 2D hippocampal surface data and the ultra-high dimensional genetic data to map the GIC pathway is urgently needed.

The literature on analysis for imaging genetics has proliferated over the past decade.
There have roughly four categories of statistical methods for the analysis.
The first is identifying genetic risk factors for scalar phenotype of interest
through genome-wide association study,
such as \cite{carrasquillo2009genetic}, \cite{bertram2012genetics} and \cite{lo2019identification}.
The second is analysis of neuroimaging data, 
ranging from acquiring raw neuroimaging data, locating brain activity, to predicting psychological, psychiatric or cognitive states \citep{lindquist2008statistical}. 
The statistical tools for detecting association between scalar phenotype of interest and imaging data include voxelwize regression \cite{zhou2014brain},
functional data analysis approach \citep{reiss2010functional},
and tensor regression models that exploits the array structure in imaging data \citep{zhou2013tensor, wang2017generalized, li2021tensor} 
The third is investigating the effects of genetic variations on imaging phenotypes.
\cite{blokland2012genetic} and \cite{zhao2019heritability} used imaging traits as phenotype and
quantify the effects of genetics on the structure and function of the human brain. 
The forth is mapping biological pathways linking genetics and imaging data to neuropsychiatric disorders and examining the joint effects of both genetic risk factors and imaging data, which remains challenging and has not been studied systematically compared to the above three categories \citep{zhu2022statistical}. Most of the exiting methods first extracted features from the imaging data and focused on the effects of the obtained features and genetic data, 
see \cite{dukart2016accurate}, \cite{ossenkoppele2021impact}, \cite{cruciani2022pls} and references therein,
which ignored the rich smoothness information in the imaging data.


To map   GIC-related pathways, we consider  a high-dimensional  Partially Functional Linear Model (PFLM) as follows: 
\begin{align} 
\label{model:PFL}
Y_i = \alpha + X_i^T\beta +	 \int_{\mathcal{T}} Z_i(t)\xi(t) dt  + \epsilon_i~~~\mbox{for}~~ i=1, \ldots, n, 
\end{align} 
where $Y_i$ is a   continuous phenotype of interest for subject $i$, $X_i\in\mathcal{X}$ is a $p\times 1$ vector of genetic and environmental variables, 
and $ Z_i(t) \in L_2(\mathcal{T}) $ is an imaging (or functional) predictor over a compact set   $\mathcal{T}$. 
Moreover, $\alpha$ is the intercept term, $\beta$ is a $p\times 1$ vector of coefficients, $\xi(t)$ is an unknown slope function, which is assumed to be in a reproducing kernel Hilbert space (RKHS) $\cH$,  and $\epsilon_i$s are measurement errors.  
We consider the case that the dimension of $\beta$ is either comparable to or much larger than the sample size $n$ and $\xi(t)$ is an  
infinite dimensional function. Our statistical problem of interest is to make statistical inference on $\beta$ and $\{\xi(t): t\in\mathcal{T}\}$. 
As illustrated in Fig \ref{fig:acyclic graph}, there exist genetic and clinical confounders which affect both hippocampal shape and behavioral deficits \citep{selkoe2016amyloid}.
Compared to the classical high dimensional linear model that only considers the genetic data,
the inclusion of imaging exposure $\xi(t)$ has two important implications for the ADNI dataset.
First,
model \eqref{model:PFL} is able to quantify the direct effects of the confounders by controlling for the imaging exposure.
Second, model \eqref{model:PFL} investigate the influence of the imaging exposure while controlling for the confounders and preserving the structure of the imaging data.

\begin{figure}
\centering
\includegraphics[width=0.8\linewidth]{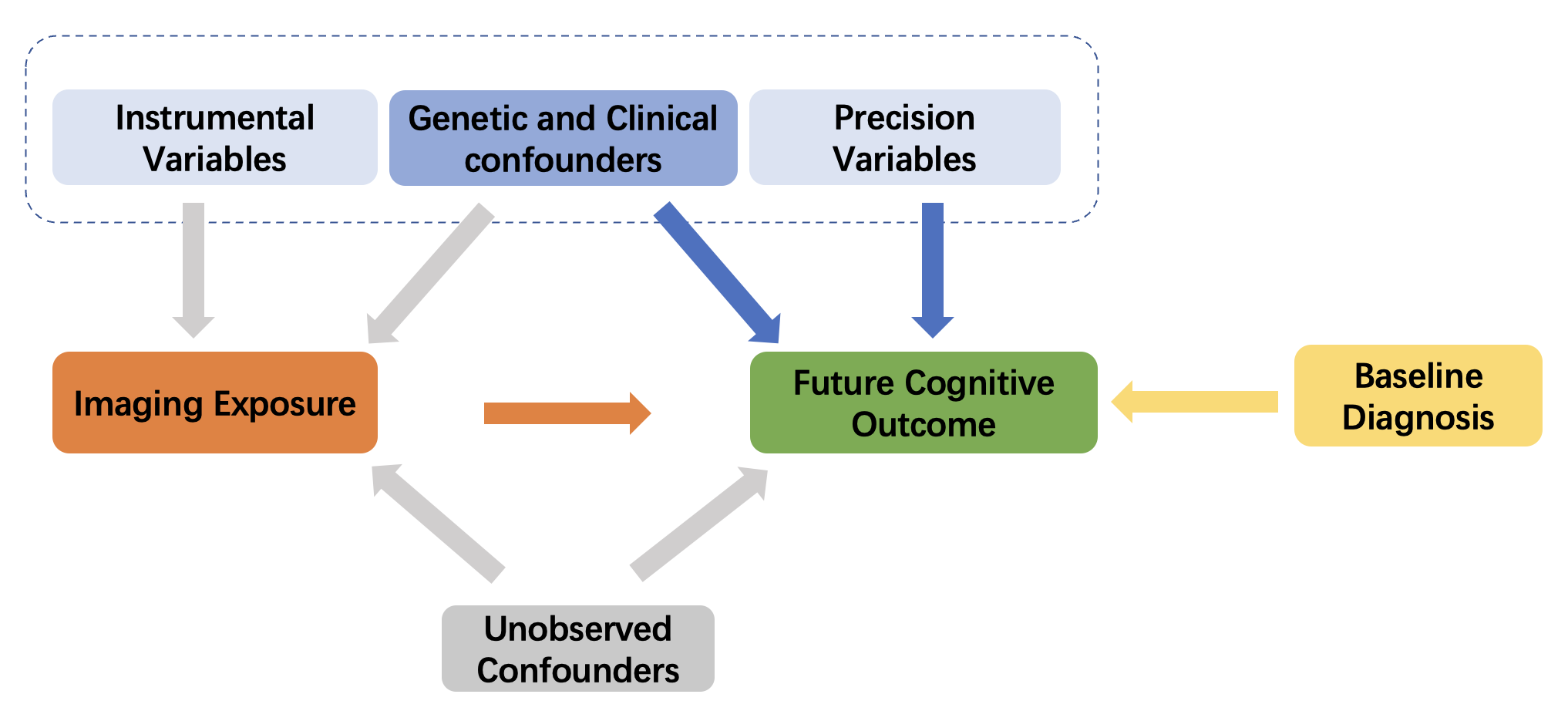}
\caption{Directed acylic graph showing potential relationships between the genetic data, the imaging data and the future outcome. The colored arrows denote the associations of interest.}
\label{fig:acyclic graph}
\end{figure}

There is  scarce literature on PFLM with high dimensional scalar covariates   with a few exceptions.
\cite{kong2016partially} studied PFLM in high dimension, in which   the dimension
of  scalar covariates was allowed to diverge with $n$.
\cite{yao2017regularized} developed a regularized partially functional quantile regression model, while  allowing 
the number of scalar predictors to increase with the sample size. 
\cite{ma2019quantile} focused on the partial functional partial regression model in ultra-high dimensions with a diverging number of 
scalar predictors.
All of the above three methods consist of three steps,    representing the functional predictors
by using their leading functional principal components (FPCs), reducing PFLM to a standard high dimensional linear regression model, 
and selecting important features through the smoothly clipped absolute deviation (SCAD) penalty \citep{fan2001variable}.
Therefore, existing approaches rely heavily on the success of the FPCA approach  \citep{wang2016functional}.

In this paper, we focus on the high dimensional PFLM  \eqref{model:PFL}, 
develop estimation method for model selection and estimation,
investigate theoretical properties of both the functional and scalar estimators, and apply the
proposed method to analyze the ADNI dataset.
We use the RKHS framework \citep{yuan2010reproducing, cai2012minimax, li2020inference} 
and impose the roughness penalty on the functional coefficient. 
The success of the existing FPCA-based methods relies on the availability of a good estimate of the functional principal components for the functional parameter,  
and may not be appropriate if the functional parameter cannot be represented effectively by the leading principals of the functional covariates \citep{yuan2010reproducing}.
On the other hand, the truncation parameter in the FPCA changes in a discrete manner, which may yield an imprecise control on the model complexity, as pointed out in \cite{ramsay2005functional}.
Furthermore, we impose the $\ell_0$ penalty on the scalar predictors due to the fact that
the $\ell_0$ penalty function is usually a desired choice among the penalty functions as it directly penalizes the cardinality of a model and seeks the most parsimonious model explaining the data.
However, it is nonconvex and the solving of an exact $\ell_0$-penalized nonconvex optimization problem involves exhaustive combinatorial best subset search, which is NP-hard and computationally challenging \citep{zhao2019simultaneous}.
We modify the computational algorithm in \cite{huang2018constructive} to deal with the above difficulty and to accommodate
the functional predictor. 
Specifically,  we proceed in three steps:  (i) profiling out the functional part by using the Representer theorem;   
(ii) simultaneously identifying the important features and    obtaining   scalar estimates; and  (iii)  plugging the scalar estimates into the loss function
to derive the functional estimate.
Meanwhile, we adapt the test statistic in \cite{li2020inference} to test the significant of the functional variable.
The implementation R code with its documentation is available as an online supplement.

Numerically, the proposed method is tested carefully on the simulated data. We also provide theoretical properties of the estimators, including the error bounds of, the asymptotic normality of the estimates of the nonzero scalar coefficients, and the null limit distribution
of the test statistic designed to test the nullity of the functional variable.
We apply PFLM to the ADNI  dataset and carry out a throughout association analysis  
between genetics, hippocampus and cognitive deficit.
Different from the existing analysis targeted to one or several cognitive measures,
the proposed method examines the joint effects of genetics and hippocampus on 13 cognitive variables
observed at 12 months after baseline measurements,
that measure different aspects of the cognitive function, and explore the shared and different heritablity patterns of the 13 cognitive scores. 
We also investigate the effect of the baseline diagnoiss information on future cognitive outcome, denoted by the yellow arrow in Fig \ref{fig:acyclic graph}.
Analysis results suggest that both the hippocampal and genetic data have heterogeneous effects on different scores.
In general, the value of both hippocampi is negatively associated with the severity of cognition impairments. Polygenic effects are observed for all the 13 cognitive scores and shared genetic etiology exists. 
The strong genetic influence is only partly attributed to
the well-known APOE4 genotype, and the baseline diagnosis status explains a larger part of the cognitive function.
There also exist strong shared genetic effect beside the effect of the APOE4 gene.
However, greater genetic heterogeneity exists within disease classifications after accounting for the baseline diagnosis status, 
These analyses 
are useful in further investigation of functional mechanisms for AD evolution.


The rest of this paper is organized as follows. 
Section \ref{sec:PFL-DataDes} includes a detailed data and problem description.
Section \ref{sec:PFL-Estimation}  describes our estimation procedure. 
Section \ref{sec:PFL-Simulation} presents  Monte Carlo simulation studies to assess 
the finite sample performance of the proposed method. 
Section \ref{sec:PFL-RealData} provides a detailed data analysis on the ADNI study.
Theoretical properties of our estimators and their proofs, additional simulation and real data analysis results can be found in the supplementary material.

\section{Data and Problem Description}
\label{sec:PFL-DataDes}
The ADNI is a large-scale multisite neuroimaging study that has collected clinical, imaging, genetic and cognitive data at multiple time points from cognitive normal (CN) subjects, subjects with mild cognitive impairment (MCI), and AD patients. It supports the investigation and development of treatments that may slow or stop the progression of AD.
The primary goal of ADNI is to test whether genetic, structural and functional neuroimaging, and clinical
data can be integrated to assess the progression of MCI and early AD.

We constructed a  data set  from the  ADNI database (adni.loni.usc.edu).
It consists of 606 subjects with 113  AD patients, 316 patients with mild cognitive impairment (MCI), and 177 normal controls (NC). 
It also includes a set of demographic variables including Age, Gender(0=Male; 1=Female), Handedness(0=Right; 1=Left), 
Retirement (0=No; 1=Yes), and Years of education.
The average age is 75.6 years old with standard deviation 6.6 years, and the average years of education is 15.7 years with standard deviation 2.9 years.
Among all the subjects, 361 are male, and 245 are female; 562 are right-handed, and 44 are left-handed; 497 are retired and 109 are not.


We extracted  GIC variables as follows.
First, we extracted 13 cognitive variables at 12 months after the onset of ADNI for measuring the severity of the cognitive impairment 
\citep{battista2017optimizing,grassi2019novel}.
See 
Table S1 in the supplementary material for 
a summary of the abbreviations of these variables.
Fig \ref{fig:corr_scores} presents the 
correlations between these scores. Among them, 
DIGITSCORE,   RAVLT.learning,  RAVLT.immediate, LDELTOTAL, and MMSE are positively correlated with lower values indicating more severe cognitive impairment,
whereas  CDRSB, FAQ, 
RAVLT.forgetting, RAVLT.perc.forgetting,  TRABSCOR, ADAS11, ADAS13, and  ADASQ4 are negatively correlated
with higher values indicating more severe cognitive impairment.  Second, we calculated the hippocampal morphometry surface measure  as a $100  \times 150$ matrix, each element of which is a continuous  variable, representing the radial distance from the corresponding coordinate on the hipppocampal surface to the medial core of the hippocampus. 
Such hippocampus surface measures  may  provide more subtle indexes compared with the volume differences in discriminating between patients with Alzheimer's and healthy control subjects \citep{li2007hippocampal}.   Third, we extracted  ultra-high dimensional genetic markers and other demongraphic covariates at baseline. 
There are $6,087,205$  
genotyped and imputed single-nucleotide polymorphisms 
(SNPs) on all of the 22 chromosomes. 

The clinical spectrum of   AD can be very  heterogeneous. 
Specifically, there are two main clinical syndromes  including
Amnestic AD with significant impairment of learning and recall and non-amnestic AD with impairment of language, visuospatial or executive function
\citep{mckhann2011diagnosis}. 
These scores measure different functions and they may lack sensitivity in different stages of AD. For example, the detection of changes of  ADAS11 and ADAS13  is limited by a substantial floor effect \citep{hobart2013putting}, whereas
CDRSB lacks sensitivity to detect changes in very early stage of AD \citep{de2021methodological}.

Little is known about the genetic  architecture of these  cognitive scores and 
the genetic-imaging-clinical (GIC) pathway for AD. 
We are particularly interested in the following scientific questions: 
\begin{itemize}
\item{(Q1)} How to quantify the joint effect of 
genetic and imaging  markers  on the thirteen cognitive scores? 

\item{(Q2)} How to measure the  shared and different  heritability patterns of the thirteen
different cognitive scores with/without correcting APOE4, which is supposed to be the stongest risk factor gene for AD?  

\item{(Q3)} How do the estimates and heritability patterns of thirteen different cognitive scores differ with/without correcting baseline diagnosis information? 

\end{itemize} 
We use  model  
\eqref{model:PFL} to  
address  (Q1)-(Q3) below. 


\begin{figure}
\centering
\includegraphics[width=0.6\linewidth]{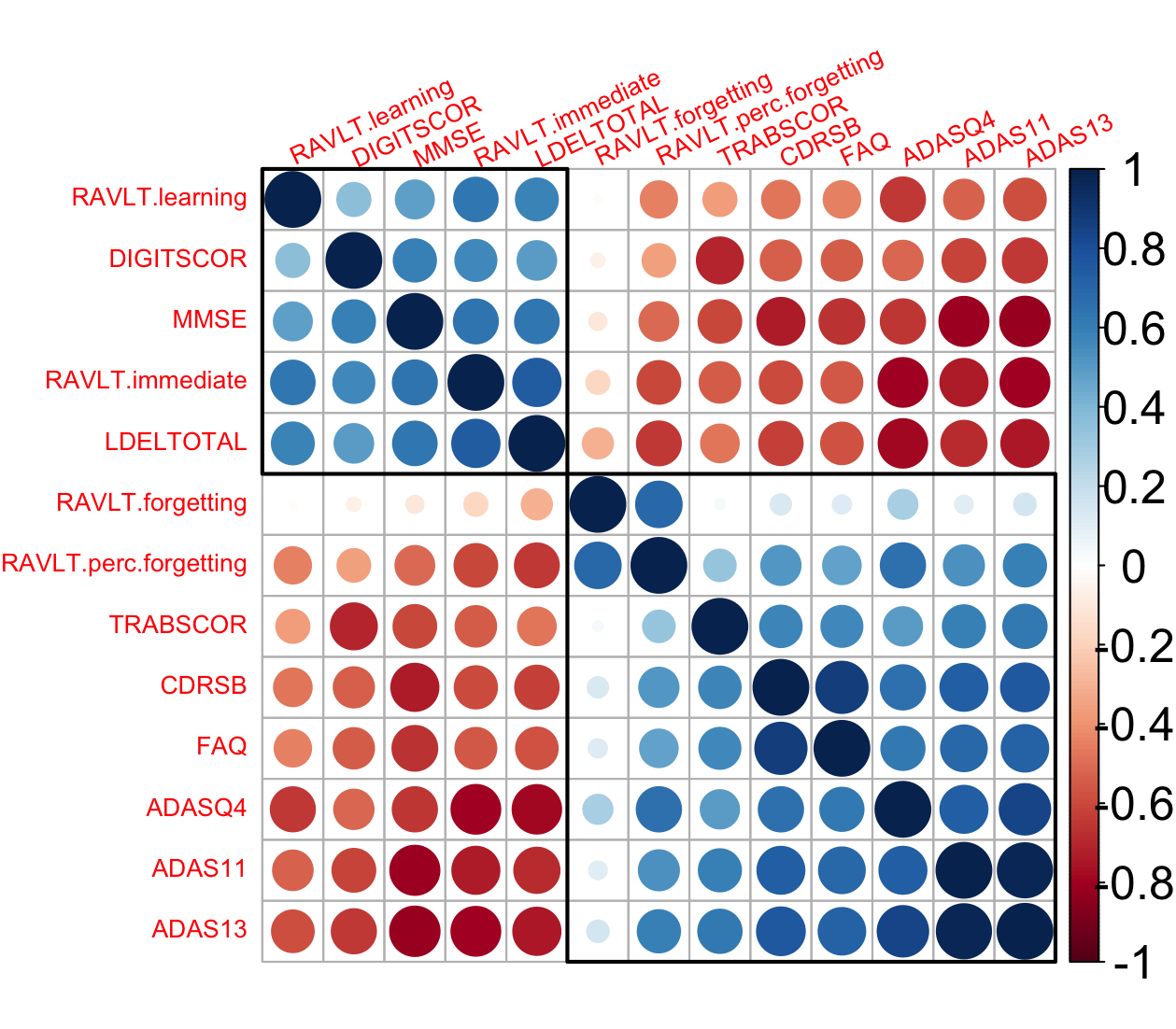}
\caption{ADNI data analysis: correlations between the 13 cognitive variables.  }
\label{fig:corr_scores}
\end{figure}

\section{Estimation and Inference Procedures} 
\label{sec:PFL-Estimation}

\subsection{Estimation Algorithm}
\label{sec:est_algorithm}
In this subsection, we develop an estimation method for  model \eqref{model:PFL}.
First, we need to introduce some notation.
Denote $ \my = (Y_1, \ldots, Y_n)^{T} $, $\mx = ( X_1^{T}, \ldots ,X_n^{T} )^{T}$, $\mb{Z}= (Z_1, \ldots, Z_n )^{T}$, 
and $ \bs{\epsilon} = ( \epsilon_1, \ldots, \epsilon_n) ^{T}  $.
Denote the true value of $\xi$ and $\beta$ as $\xi^*$ and $\beta^*$, respectively.
Let $S = \{1, 2, \ldots, p\}$.
For any $A$ and $B \subseteq S$ with length $| A |$ and $| B |$, denote $\beta_A =( \beta_i, i \in A ) \in \mathbb{R}^{ |A| }$.
Denote $\beta |_A \in \mathbb{R}^p$ be a vector with its $i$-th element $( \beta |_A )_i = \beta_i 1 (i \in A)$, where
$1(\cdot)$ is the indicator function. 
Let $\| \beta \|_{k, \infty}$ be the $k$th largest elements in absolute value.
Denote $\| \cdot \|_0$ as the $\ell_0$ norm that calculates the number of nonzero elements of a vector.
Let $\| \cdot \|_2$ be the $\ell_2$-norm such that $ \| \beta\|^2_2 =  \sum_{i=1}^p \beta_i^2 $ and 
$\| \cdot \|_{L_2}$ be the $L_2$-norm such that $\|  \xi  \|^2_{L_2} =\int_{\mathcal{T}} \xi^2 (t) dt $.
Thus, we assume throughout $E(X)=0$, $E\{Z(t)\}=0$ and $E(Y)=0$, and therefore the intercept term can be ignored.
In practice, we assume that the response $Y_i$ and the predictors $X_i$ and $Z_i(\cdot)$ are all mean centered such that $n^{-1} \sum_{i=1}^n Y_i = 0 $, $n^{-1}\sum_{i=1}^n X_i = \mb{0}$, and $ n^{-1}\sum_{i=1}^n Z_i(\cdot) = 0$. 

We use the least-squares loss to estimate the functional and scalar coefficients.
Due to the infinite-dimensional functional coefficient and high dimensional scalar coefficients, regularizations are needed for estimating both
$\xi(t)$ and $\beta$. 
Similar to \cite{yuan2010reproducing} and \cite{cai2011optimal}, the functional coefficient $\xi^*$ is assumed to reside in a RKHS
$\mathcal{H}(K)$ with a reproducing kernel $K$. The RKHS roughness penalty is imposed on the functional parameter, while
the $\ell_0$ penalty is imposed on the scalar parameters following a similar spirit of \cite{huang2018constructive}.
Therefore, we   solve the following minimization problem
\begin{align*}
\min_{\beta\in\mathbb{R}^p, \xi \in \cH }  (2n)^{-1} \sum_{i=1}^n \left[ Y_i - \left( X_i^T\beta + \int_\mathcal{T} Z_i(t)\xi(t) \mathrm{d}t  \right) \right]^2 \quad \text{subject to }  \|\beta\|_0 \leq J, ~ \| \xi \|_{\cH}^2 \leq \widetilde J,
\end{align*}
where $J >0$ controls the sparsity level of $\beta$  and $\widetilde J >0$ controls the smoothness level of $\xi$.
Consider the Lagrangian form of the above minimization problem
\begin{align}
\min_{\beta\in\mathbb{R}^p, \xi\in \cH } \Biggr\{ (2n)^{-1} \sum_{i=1}^n \left[ Y_i - \left( X_i^T\beta + \int_\mathcal{T} Z_i(t)\xi(t) \mathrm{d}t  \right) \right]^2
+ \tau\|\beta\|_0 + {0.5 \lambda }\| \xi \|_{\cH}^2  \Biggr\}, \label{eq:min}
\end{align} 
where $\tau$ and $\lambda$ are the Lagrange multipliers. 
To solve the minimization problem (\ref{eq:min}), the following Representer theorem is very useful.
\begin{theorem}
\label{thm:representer theorem}	
For any $\beta \in\mathbb{R}^p$, there exists a parameter vector $\mb{c}({\beta})$ such that
\begin{align}
	\widehat{{\xi}}({\beta}) = \sum_{i=1}^n c_{i}(\beta)  (K Z_{i} ) \label{eq:PFL-representer}
\end{align}
where 
$c_{i}(\beta)$ is the $i$-th component of $\mb{c}({\beta})$ and 
$K f := \int_{\mathcal{S}} K(\cdot,t)f(t)\mathrm{d}t $ for   $f\in L_2 ( \mathcal{T})$.
\end{theorem}

Define $
\bs{\Sigma}_{ii'} =  \iint_{\mathcal{S}\times\mathcal{S}} Z_{i}(s) K(s,t) Z_{i'}(t)\mathrm{d}s\mathrm{d}t
$ as the  $(i,i')$-th entry of  $\bs{\Sigma}$. 
The objective function in (\ref{eq:min}) can be written in matrix form as
\begin{align}
(2n)^{-1} \left\| \my - \mx\beta - \bs{\Sigma} \mb{c}  \right\|_2^2 + \tau\|\beta\|_0 +{0.5\lambda }\mb{c}^T\bs{\Sigma}\mb{c}.  \label{eq:min3}
\end{align}
Taking the first-order derivative of (\ref{eq:min3}) with respect to $\mb{c}$ and setting it  to zero give
\begin{align} \label{eq:coef_xi}
\mb{c} = (\bs{\Sigma}+n\lambda\mi)^{-1} (\my-\mx\beta).
\end{align}
Substituting (\ref{eq:coef_xi}) into (\ref{eq:min3}), we obtain the following minimization problem
\begin{align}
\min_{\beta\in\mathbb{R}^p} \left\{  (2n)^{-1} (\my-\mx\beta)^T\mb{P}_\lambda(\my-\mx\beta) + \tau\|\beta\|_0 \right\}, \label{eq:min4}
\end{align}
where $ \mb{P}_\lambda := n\lambda (\bs{\Sigma} +n\lambda\mi )^{-1} $. Once we find an approximate solution to \eqref{eq:min4}, we can plug it back into \eqref{eq:coef_xi}, leading to an estimate of $\xi$.

We derive the KKT conditions for  \eqref{eq:min4}. If $\beta^o$ is a minimizer of \eqref{eq:min4}, then we have
\begin{align} \label{eq:kkt}
d^o = \mx^T\mb{P}_\lambda(\my - \mx\beta^o)/n \quad \text{and} \quad \beta^o = H_\tau(\beta^o+d^o),
\end{align}
where $H_\tau(\cdot)$ is the elementwise hard thresholding operator with its $i$-th entry  defined by
$
H_\tau(\beta)_i =  
0,$ if $ |\beta_i| < \sqrt{2\tau}$ and 
$\beta_i,$  if $|\beta_i| \geq \sqrt{2\tau}. $
Conversely, if $\beta^o$ and $d^o$ satisfy \eqref{eq:kkt}, then $\beta^o$ is a local minimizer of \eqref{eq:min4}.

Let $A^o = \{ i : \beta_i \neq 0 \}$ and $I^o=\{i:\beta_i=0\}$. 
Denote $\beta_{A^o}^o = ( \beta_i : i \in A^o )\in \mathbb{R}^{|A^o|}$ and similarly $\beta_{I^o}^o$, $d_{A^o}^o$, and $d_{I^o}^o$. Denote $\mx_{A^o} = (\mx_i:i\in A^o) \in \mathbb{R}^{n\times |A^o|}$ and similarly $\mx_{I^o}$. By \eqref{eq:kkt}, we have
$
A^o = \{ i : |\beta_i^o + d_i^o| \geq \sqrt{2\tau} \},$  $  I^o = \{ i  :  |\beta_i^o+d_i^o| < \sqrt{2\tau} \},
$ 
and the system of equations
\begin{equation*}
\beta_{I^o}^o = \bs{0},~~~   
d_{A^o}^o  = \bs{0},  
\beta_{A^o}^o = (\mx_{A^o}^T\mb{P}_\lambda \mx_{A^o} )^{-1}\mx_{A^o}^T\mb{P}_\lambda\my,~~~
d_{I^o}^o = \mx_{I^o}^T\mb{P}_\lambda (\my-\mx_{A^o}\beta_{A^o}^o)/n.
\end{equation*}
If we want $\beta^o$ to have exactly $J$ nonzero elements, then we can set $\sqrt{2\tau}$ equal to the $J$-th
largest element of the sequence $ \{ |\beta_i^o + d_i^o| : i = 1, \ldots , p\}$.

To solve the above system of equations and obtain both the functional and scalar estimates,
for a given sparsity level $J$, we modify the support detection and root finding algorithm
\citep{huang2018constructive} to accommodate the functional variable.
We use the Generalized Cross Validation (GCV) to select the tuning parameter $\lambda$ by  following the same reasoning in \cite{yuan2010reproducing}. To select an appropriate sparsity level $J$, we use   the
high dimensional Bayesian information criterion (HBIC)
\citep{wang2013calibrating} as follows: 
\begin{align*}
\text{HBIC}_J =\log \left(  {n}^{-1} \| \my - \mx\widehat{\beta} - \int_{\mathcal{T}} \mb{Z}(t)\widehat{\xi}(t) dt\|^2 \right) + J\log \log(n) \log(p)/n.
\end{align*}
The sparsity level $J$ is set to minimize HBIC$_J$.

We summarize the above algorithm  in Algorithm \ref{alg:estimation}.
The proposed estimation algorithm can be easily extended to allow for certain sets of covariates not to be penalized by including these covariates in the active set at each iteration.

\begin{algorithm}[h] 
\caption{Functional support detection and root finding (FSDAR)} 
\label{alg:estimation}
\begin{algorithmic}[1] 	
	\Require 
	An initial $\beta^0$ and the sparsity level $J$; set $k=0$.
	\State 
	for a given $\lambda$,
	calculate $d^0 = \mx^T\mb{P}_{\lambda}(\my-\mx\beta^0)/n$ and  $\mb{P}_{\lambda} =n\lambda(\bs{\Sigma}+n\lambda\mi)^{-1} $; 
	\For{$k=0,1,2,\ldots$}
	\State $A^k = \{ i: |\beta_i^k+d_i^k| \geq \|\beta^k + d^k\|_{J,\infty} \}, ~~ I^k = (A^k)^c$;
	\State $\beta_{A^k}^{k+1} = (\mx_{A^k}^T\mb{P}_{\lambda}\mx_{A^k} )^{-1} \mx_{A^k}^T\mb{P}_{\lambda}\my, ~~  \beta_{I^k}^{k+1} = \bs{0}$;
	\State  $d_{A^k}^{k+1} = \bs{0}, ~~ d_{I^k}^{k+1} = \mx_{I^k}^T\mb{P}_{\lambda^k}(\my-\mx_{A^k}\beta_{A^k}^{k+1})/n$;
	\If{$A^{k+1} =A^k$}
	\State Stop and denote $\widehat{\beta} = (  \widehat \beta_{A^k}^{T}, \widehat \beta_{I^k}^{T}  )^{T}$.
	\Else
	\State $k=k+1$; 
	\EndIf		
	\EndFor
	\State
	for a list of candidate $\lambda$s, repeat steps 1-11 and select 
	$\lambda$ that gives the best GCV$=n\| \mb{P}_\lambda(\my-\mx_{A^k}\beta^k_{A^k}) \|_2^2 / [\text{tr}(\mb{P}_\lambda)]^2$
	\Ensure 
	$\widehat{\beta},~ \widehat{ \mb{c} }= (\bs{\Sigma}+n\lambda \mi)^{-1} (\my-\mx \widehat{\beta}), ~~\mbox{and}~~
	~\widehat{{\xi}} = \sum_{i=1}^n \widehat{ \mb{c} }_i(KZ_{i} )$. 
\end{algorithmic} 
\end{algorithm}

\subsection{Computational Complexity}
\label{sec:computational_complexity}
We discuss the computational complexity of Algorithm \ref{alg:estimation} as follows. Denote $m$ the number the of points 
of the functional variable.
It takes $O(nm^2 + n^2 m )$ to calculate the matrix $\bs{\Sigma}$, which can be precomputed and stored.
The computational complexity of calculating the matrix $\mb{P}_{\lambda}$ is $O(nm^2 + n^2 m +n^3 )$ and that of calculating $d^0$ is $O(n^2 p)$.
Hence, the complexity of step 1 is $O(n^2 p)$ under the condition that $p$ is much larger than the sample size
It takes $O(p)$ flops for step 3, $O(n^2 J)$ for step 4 and step 5,
and $O(p)$ for checking the stopping condition in step 6.
As a result, for a given $\lambda$ and the sparsity level $J$, the overall cost per iteration of
Algorithm is $O(n^2 p)$.
It needs no more than $O(\log(R))$ iterations to get a good solution in Corollary 2 of the supplementary material, where $R = \max \{ | \beta_i^*  |,  i \in A^*\} / \min \{ | \beta_i^*  |,  i \in A^*\}$.
Then the overall cost of Algorithm \ref{alg:estimation} is $O(n^2 p \log(R))$.

If we choose the tuning parameters $\{\lambda, J \}$ by a grid search method, then the computational complexity increases by a factor of the number of the set of tuning parameters.

\subsection{Inference for the Functional Predictor}
Except for the estimation,
it is also important to test the nullity of the functional covariate $\xi(t)$.
For instance, in our real data analysis, we are interested in examining whether the hippocampus data have significant effects on the cognitive decline. Therefore, we propose to test
\begin{align} 
\label{eq:test_functional}
H_0: \xi_0(t)=0, ~ \text{for any }t, \quad \text{vs.}~
H_1: \xi_0(t) \neq 0, ~ \text{for some }t.
\end{align} 

After variable selection, we adapt the test statistic in \cite{li2020inference} to test
\eqref{eq:test_functional}. Specifically, 
$$
T_\xi = 2n [ \ell_{n\lambda }( \hat \beta_{H_0}, \xi_0(t)) - \ell_{n\lambda }( \hat \beta, \hat \xi(t)) ],
$$
where $ \ell_{n\lambda }$ is the loss function and $\hat \beta_{H_0}$ is the estimator under the null hypothesis. 
Following \cite{li2020inference}, we can show that $T_\xi$ converges to a normal distribution and approximate a chi-square distribution in Corollary 4 of the supplementary material.

For theoretical guarantee, We establish the theoretical properties of the estimators, including the non-asymptotic error bounds of $\widehat \beta_k$ at each iteration, the general non-asymptotic error bound of $\widehat \xi(t)$, the asymptotic normality of the estimates of the nonzero scalar coefficients and the null limit distribution of $T_\xi$ in the supplementary material.

\section{Simulation Studies}
\label{sec:PFL-Simulation}

We examine the finite sample performance of the proposed estimation method in two cases, including one dimensional   $\xi(s)$ in this section and two dimensional $\xi(s)$ 
in the supplement material. We apply algorithm \ref{alg:estimation} to estimate the unknown coefficients. The initial value is also set to be zero and we choose $J \in \{1, 2, \dots, 50 \}$ and $\lambda$ from 50 evenly spaced points on  $[1e-5, 0.1]$.

\begin{example}
\label{ex:comparisions with yao}
The following is designed to evaluate the estimation and prediction performances for one dimensional $\xi(t)$ in 
$\mathcal{T}=[0,1]$.  The functional predictor $Z(t)$ is of the form
$
Z(t) = \sum_{k=1}^{50}  U_k \phi_k (t)$ for  $t \in [0,1], 
$
where $\phi_{2l-1} (t) = \sqrt{2} \cos( (2l-1) \pi t )$ and $\phi_{2l} (t) = \sqrt{2} \sin( (2l-1) \pi t ), {\l=1, \ldots, 25}$, and 
$\{ U_k \}$ are independently sampled from the normal distribution $ N(0, 16 | k - C_0  |+1) $ with $C_0 \in \{ 1,3\}$.
For the coefficient function, we set
$\xi (t) = \sum_{k=1}^{50} 4 (-1)^{k+1} k^{-2} \phi_k(t).
$
When $C_0 =1$, the functional coefficient can be efficiently represented in terms of the leading functional principal components.
When $C_0 =3$, the representative basis functions for $Z(t)$ and $\xi(t)$ are disordered such that
the leading eigen-functions $\phi_k(t)$ of the covariance kernel of $Z(t)$ are around $k=3$.

Following \cite{kong2016partially}, we allow moderate correlation between $Z(t)$ and the scalar covariates $X = (X_1, \ldots, X_p)^{T} $
by introducing  a correlation structure between 
$\{ U_1, U_2, U_3, U_4 \}$ and $X = (X_1, \ldots, X_p)^{T} $ as 
corr$( U_k, X_l ) = \rho_1^{ |k-l| +1 }$ for $k=1, \ldots, 4$ and $l=1,  \ldots,  p$ with $\rho_1 \in \{0.2, 0.4\}$.
The scalar covariates $X = (X_1, \ldots, X_p)^{T} $ are jointly normal with zero mean, unit variance, and AR($\rho_2$)
with $\rho_2 \in \{0.3, 0.5, 0.7\}$.  
For each subject $i$, we observe $Z_i(t_{ij})$ at 100 equally spaced points. 
The errors $\epsilon_i$s are generated from the standard normal distribution.
The sample size is chosen to be $n=200$.
We consider $\beta$ with two different values of dimensionality: $p=150$, which is \emph{smaller} than the training sample size, and $p=1500$, which is \emph{larger} than the training sample size. Specifically, the underlying true $\beta$ is set to be
$
\beta = (3,1.5,1,2.5,2,\underbrace{0,\ldots,0}_{p-5})^T.
$   
Besides the proposed method, the method based on FPCA proposed by \cite{kong2016partially} is also considered for comparison.
The number of the functional components and the penalty tuning parameter are selected by minimizing the value of 
HBIC. 
The initial $\beta^0$ is set to be 0. We have tried several different choices, including $\beta^0=(1, \dots, 1)^\top$, $\beta^0=(10, \dots, 10)^\top$,
$\beta^0=(100, \dots, 100)^\top$ and the marginal correlation between the covariates and the response.
The results are the same, which are not sensitive to the selection of the initial $\beta^0$.

All simulation results are based on 200 replications by using R (version 3.6.0) on a Linux server
(equipped with Intel(R) Xeon(R) CPU E5-2640 v4 @ 2.40GHz, 125 GB RAM).
We evaluate the estimation accuracy of   $\widehat{\beta}$   by using the mean squared error
MSE$_{\beta} = \| \widehat{\beta}-\beta \|_2^2$ and 
that of $\xi$
by using the mean integrated squared error MSE$_{\xi}= \| \widehat{\xi}-\xi \|_{L_2}^2 $ as well as 
the relative MSE of  $\widehat{\xi}$ such that RMSE$_\xi= \| \widehat{\xi}-\xi \|_{L_2}^2 / \| \xi \|_{L_2}^2 $. We also calculate 
the number of false zero scalar predictors (FZ), the number of false nonzero scalar predictors (FN), and
the prediction mean squared error (PMSE) based on 200 new test samples. 
We further calculate the compatation time (in seconds).

Table \ref{table:low_dimension} presents the variable selection accuracy, estimation accuracy, and prediction results for
the moderate number of scalar variables with $n=200$ and $p=150$.
Our  method outperforms the method based on FPCA in  \cite{kong2016partially} in almost all scenarios.
Specifically, the selection of scalar predictors for our  method is more accurate and more stable than the competing method 
with smaller numbers of false nonzero scalars and zero false zero scalars.  For our method, x
the number of false zero scalars and that of  false nonzero scalars  do not differ too much across different correlations among the scalar variables.
However, FZ and FN of the competing method  \citep{kong2016partially} 
increase as the correlation among the scalar variables becomes larger. It indicates that 
more zero scalar variables would be selected in PFLM, whereas more nonzero scalar variables would be excluded from PFLM.
When the representative basis functions for $Z(t)$ and $\xi(t)$ are not exactly matched, our method still yields stabler estimates than  
the competing method  \citep{kong2016partially}.
Furthermore, MSEs  and PMSEs for our method are smaller than 
those for the competing method  \citep{kong2016partially} in all scenarios.

Table \ref{table:high_dimension} reports additional simulation results corresponding to   $n=200$ and $p=1500$.
The proposed method outperforms the competing method \citep{kong2016partially} 
in terms of FNs, FZs,  MSEs, and PMSEs. For instance, 
it is noteworthy that the number of false zero scalars for the competing method \cite{kong2016partially}
increases as  $\rho_2$ increases.

\begin{table}
	\begin{center}
		\caption{ Simulation results of Monte Carlo averages and empirical  standard errors in parentheses for $n=200, p=150$.}
		\label{table:low_dimension}
		
		\vspace{1ex}
		\scriptsize
		\tabcolsep 3pt
		\begin{tabular}{cccccccccccc}
			\hline
			\hline
			center	&	$\rho_1$  &      $\rho_2$    &      &   FZ    &  FN  &  MSE$_{\beta}$    &   MSE$_{\xi}$   &   RMSE$_{\xi}$    &   PMSE  & Time(s)   \\
			\hline
			1          &   0.2    &  0.3  &   Proposed      &    0.000(0.000) &  0.670(0.737) &  0.067(0.046) &  0.035(0.023) &  0.002(0.001) &  1.085(0.121) & 2.219(2.458)  \\
			&           &         &   FPCA              &  0.005(0.071) & 	1.115(2.875) & 	0.152(0.307) & 	0.103(0.051) & 	0.006(0.003) & 	1.269(0.273) & 0.590(0.098)  \\
			\cline{3-11}
			&            &  0.5  &   Proposed         &  0.000(0.000) &  0.685(0.767) &  0.082(0.056) &  0.036(0.026) &  0.002(0.001) &  1.087(0.122)   & 2.714(2.987) \\
			&           &      &   FPCA                   &   0.015(0.122) & 	1.390(2.890) & 	0.224(0.346) & 	0.103(0.051) & 	0.006(0.003) & 	1.271(0.228)   & 0.694(0.122) \\
			\cline{3-11}
			&            &  0.7  &   Proposed         &  0.000(0.000) &  0.530(0.679) &  0.107(0.072) &  0.036(0.024) &  0.002(0.001) &  1.076(0.123)   & 2.737(2.744) \\
			&           &      &   FPCA                  &   0.255(0.437) & 	1.350(3.015) & 	0.779(1.874) & 	0.110(0.055) & 	0.006(0.003) & 	1.377(1.083)  & 0.685(0.125) \\
			\cline{2-11}	
			&   0.4    &  0.3  &   Proposed         &  0.000(0.000) &  0.695(0.731) &  0.072(0.046) &  0.039(0.029) &  0.002(0.002) &  1.081(0.114)   & 2.572(2.537) \\
			&           &      &   FPCA               &   0.015(0.122) & 	1.150(2.594) & 	0.183(0.367) & 	0.115(0.065) & 	0.007(0.004) & 	1.284(0.349)   & 0.701(0.123) \\
			\cline{3-11}
			&            &  0.5  &   Proposed         &  0.000(0.000) &  0.635(0.703) &  0.079(0.056) &  0.038(0.032) &  0.002(0.002) &  1.083(0.115)  & 2.981(2.930)  \\
			&           &      &   FPCA               &    0.035(0.184) & 	1.155(2.448) & 	0.304(0.523) & 	0.113(0.062) & 	0.006(0.004) & 	1.302(0.306)  & 0.678(0.120)  \\
			\cline{3-11}
			&            &  0.7  &   Proposed         &   0.000(0.000) &  0.510(0.680) &  0.111(0.079) &  0.038(0.024) &  0.002(0.001) &  1.078(0.118)  & 2.423(2.518) \\
			&           &      &   FPCA                   &   0.300(0.470) & 	1.020(2.143) & 	0.658(1.016) & 	0.122(0.066) & 	0.007(0.004) & 	1.310(0.315)  & 0.704(0.125)  \\	 
			\hline		               	               
			3           &    0.2     &   0.3  &   Proposed         &   0.000(0.000) &  0.655(0.706) &  0.067(0.046) &  0.026(0.016) &  0.001(0.001) &  1.088(0.118)  & 2.527(2.441) \\
			&           &         &   FPCA                &   0.000(0.000) & 	0.125(0.448) & 	0.052(0.076) & 	0.193(0.103) & 	0.011(0.006) & 	1.369(0.185) & 0.712(0.134) \\ 
			\cline{3-11}
			&             &    0.5  &   Proposed          &  0.000(0.000) &  0.665(0.752) &  0.081(0.055) &  0.026(0.017) &  0.001(0.001) &  1.086(0.123)  & 2.338(2.231) \\
			&           &      &   FPCA               &  0.000(0.000) & 	0.250(0.714) & 	0.065(0.067) & 	0.191(0.102) & 	0.011(0.006) & 	1.365(0.178) & 0.702(0.140) \\
			\cline{3-11}
			&             &    0.7  &   Proposed     &  0.000(0.000) &  0.520(0.687) &  0.105(0.073) &  0.025(0.015) &  0.001(0.001) &  1.080(0.124)   & 2.765(2.791) \\
			&           &      &   FPCA               &    0.050(0.218) & 	0.340(0.805) & 	0.196(0.377) & 	0.187(0.095) & 	0.011(0.005) & 	1.369(0.190) & 0.683(0.122) \\
			\cline{2-11}	
			&   0.4    &  0.3  &   Proposed         &  0.000(0.000) &  0.650(0.728) &  0.068(0.045) &  0.028(0.019) &  0.002(0.001) &  1.081(0.118)  & 2.396(2.307)  \\
			&           &      &   FPCA               &  0.000(0.000) & 	0.090(0.335) & 	0.053(0.049) & 	0.190(0.106) & 	0.011(0.006) & 	1.361(0.174)  & 0.667(0.118)   \\
			\cline{3-11}
			&            &  0.5  &   Proposed          &  0.000(0.000) &  0.555(0.692) &  0.072(0.055) &  0.027(0.017) &  0.002(0.001) &  1.080(0.113)  & 2.637(2.744) \\
			&           &      &   FPCA               &     0.000(0.000) & 	0.195(0.573) & 	0.073(0.079) & 	0.197(0.099) & 	0.011(0.006) & 	1.368(0.163)  & 0.668(0.124) \\
			\cline{3-11}
			&            &  0.7  &   Proposed          &  0.000(0.000) &  0.530(0.715) &  0.107(0.082) &  0.027(0.015) &  0.002(0.001) &  1.081(0.118) & 2.795(2.912)   \\
			&           &      &   FPCA               &     0.110(0.314) & 	0.225(0.553) & 	0.245(0.456) & 	0.201(0.125) & 	0.011(0.007) & 	1.389(0.209) & 0.664(0.120)  \\	 
			\hline \hline
		\end{tabular}
	\end{center}
\end{table}

\begin{table}
	\begin{center}
		\caption{ Simulation results of Monte Carlo averages and empirical standard errors in parentheses for $n=200, p=1500$.}
		\label{table:high_dimension}
		
		\scriptsize
		\tabcolsep 3pt
		\begin{tabular}{cccccccccccc}
			\hline
			\hline
			center	&	$\rho_1$  &      $\rho_2$    &      &   FZ    &  FN  &    MSE$_{\beta}$    &   MSE$_{\xi}$   &   RMSE$_{\xi}$    &   PMSE  & Time(s)  \\
			\hline
			1	&	0.2	&	0.3	&	Proposed	&	0.000(0.000) & 	0.820(0.813) & 	0.098(0.068) & 	0.094(0.011) & 	0.005(0.001) & 	1.191(0.135) & 	35.566(33.777) 	\\
			&		&		&	FPCA	&	0.005(0.071) & 	4.655(9.863) & 	0.224(0.763) & 	0.107(0.078) & 	0.006(0.004) & 	1.344(0.812) & 	1.079(0.233) 	\\
			\cline{3-11}															
			&		&	0.5	&	Proposed	&	0.000(0.000) & 	0.855(0.811) & 	0.121(0.075) & 	0.094(0.012) & 	0.005(0.001) & 	1.207(0.143) & 	38.249(34.465) 	\\
			&		&		&	FPCA	&	0.035(0.184) & 	5.190(9.596) & 	0.429(1.111) & 	0.111(0.057) & 	0.006(0.003) & 	1.418(0.964) & 	1.090(0.250) 	\\
			\cline{3-11}															
			&		&	0.7	&	Proposed	&	0.000(0.000) & 	0.665(0.778) & 	0.150(0.098) & 	0.092(0.010) & 	0.005(0.001) & 	1.178(0.137) & 	38.350(34.027) 	\\
			&		&		&	FPCA	&	0.410(0.493) & 	5.480(9.252) & 	1.031(1.295) & 	0.115(0.059) & 	0.007(0.003) & 	1.465(0.760) & 	1.094(0.219) 	\\
			\cline{2-11}															
			&	0.4	&	0.3	&	Proposed	&	0.000(0.000) & 	0.715(0.766) & 	0.108(0.065) & 	0.100(0.013) & 	0.006(0.001) & 	1.183(0.138) & 	39.534(34.614) 	\\
			&		&		&	FPCA	&	0.010(0.100) & 	4.560(9.637) & 	0.233(0.616) & 	0.118(0.067) & 	0.007(0.004) & 	1.346(0.620) & 	1.047(0.247)  	\\
			\cline{3-11}															
			&		&	0.5	&	Proposed	&	0.000(0.000) & 	0.675(0.736) & 	0.121(0.075) & 	0.097(0.012) & 	0.005(0.001) & 	1.174(0.133) & 	37.203(32.144)  	\\
			&		&		&	FPCA	&	0.065(0.247) & 	5.205(8.870) & 	0.501(1.111) & 	0.125(0.063) & 	0.007(0.004) & 	1.425(0.783) & 	1.096(0.236)  	\\
			\cline{3-11}															
			&		&	0.7	&	Proposed	&	0.000(0.000) & 	0.685(0.767) & 	0.179(0.113) & 	0.095(0.011) & 	0.005(0.001) & 	1.173(0.143) & 	35.889(31.236)  	\\
			&		&		&	FPCA	&	0.455(0.509) & 	4.795(8.094) & 	1.089(1.434) & 	0.139(0.086) & 	0.008(0.005) & 	1.440(0.562) & 	1.090(0.223)  	\\
			\hline															
			3	&	0.2	&	0.3	&	Proposed	&	0.000(0.000) & 	0.205(0.473) & 	0.081(0.068) & 	0.073(0.017) & 	0.004(0.001) & 	1.467(0.185) & 	36.373(33.003) 	\\
			&		&		&	FPCA	&	0.000(0.000) & 	0.395(0.961) & 	0.057(0.059) & 	0.217(0.135) & 	0.012(0.008) & 	1.391(0.178) & 	1.045(0.235)  	\\
			\cline{3-11}															
			&		&	0.5	&	Proposed	&	0.000(0.000) & 	0.240(0.494) & 	0.108(0.076) & 	0.072(0.016) & 	0.004(0.001) & 	1.471(0.188) & 	39.177(32.842)  	\\
			&		&		&	FPCA	&	0.005(0.071) & 	0.940(1.565) & 	0.103(0.201) & 	0.224(0.134) & 	0.013(0.008) & 	1.413(0.196) & 	1.078(0.270)  	\\
			\cline{3-11}															
			&		&	0.7	&	Proposed	&	0.000(0.000) & 	0.170(0.427) & 	0.169(0.132) & 	0.071(0.018) & 	0.004(0.001) & 	1.459(0.181) & 	37.563(33.131)  	\\
			&		&		&	FPCA	&	0.130(0.337) & 	1.910(2.755) & 	0.274(0.484) & 	0.219(0.127) & 	0.012(0.007) & 	1.416(0.216) & 	1.079(0.203) 	\\
			\cline{2-11}															
			&	0.4	&	0.3	&	Proposed	&	0.000(0.000) & 	0.255(0.549) & 	0.146(0.081) & 	0.088(0.021) & 	0.005(0.001) & 	1.458(0.183) & 	34.386(31.798)  	\\
			&		&		&	FPCA	&	0.000(0.000) & 	0.485(1.080) & 	0.072(0.072) & 	0.222(0.139) & 	0.013(0.008) & 	1.402(0.185) & 	1.078(0.240)  	\\
			\cline{3-11}															
			&		&	0.5	&	Proposed	&	0.000(0.000) & 	0.235(0.540) & 	0.164(0.100) & 	0.082(0.018) & 	0.005(0.001) & 	1.452(0.182) & 	38.020(32.011) 	\\
			&		&		&	FPCA	&	0.010(0.100) & 	0.905(1.568) & 	0.133(0.229) & 	0.227(0.137) & 	0.013(0.008) & 	1.420(0.204) & 	1.061(0.228)  	\\
			\cline{3-11}															
			&		&	0.7	&	Proposed	&	0.000(0.000) & 	0.205(0.463) & 	0.277(0.166) & 	0.078(0.018) & 	0.004(0.001) & 	1.448(0.182) & 	39.756(33.367)  	\\
			&		&		&	FPCA	&	0.195(0.397) & 	1.685(2.270) & 	0.388(0.600) & 	0.236(0.137) & 	0.013(0.008) & 	1.434(0.221) & 	1.083(0.223)  	\\	 
			\hline \hline	               
		\end{tabular}
	\end{center}
\end{table}

Fig \ref{fig:solpath} shows the solution path of $\widehat{\beta}$ for different $\rho_2$ corresponding to $(p, C_0, \rho_1)=(150, 1, 0.2)$. The solution path displays how $\widehat{\beta}$ evolves either as  the sparsity level of the proposed method increases or as the penalty tuning parameter decreases.
Specifically, the colored lines show how $\widehat{\beta_1}, \ldots, \widehat{\beta_5}$  changes with the sparsity level and the penalty tuning parameter.
For our method,
when $\rho_2$ is small (e.g., $\rho_2=0.3$), the five variables gradually enter the model and the more significant the variable is, the earlier it is  selected. However, 
when $\rho_2$ increases to a larger value,  the less important variable may enter the model earlier than the more important one. 
The evolution   of significant $\widehat{\beta}$s does not change very much when the sparsity level is greater than the true sparsity level 5. 
For the competing method  \citep{kong2016partially}, we observed similar entering orders of the variables.
The values of significant $\widehat{\beta}$s also do not differ very much when the tuning parameter is smaller than some value when $\rho_2$ is small.
However, when the correlation among the scalar variables increases, the evolution of significant $\widehat{\beta}$s shows different patterns such that nonzero scalar variables may be excluded from PFLM  as the tuning parameter of the SCAD penalty decreases.
These results may indicate that our method is more stable and  accurate than the competing method.

\begin{figure}
	\centering
	\includegraphics[width=0.9\linewidth]{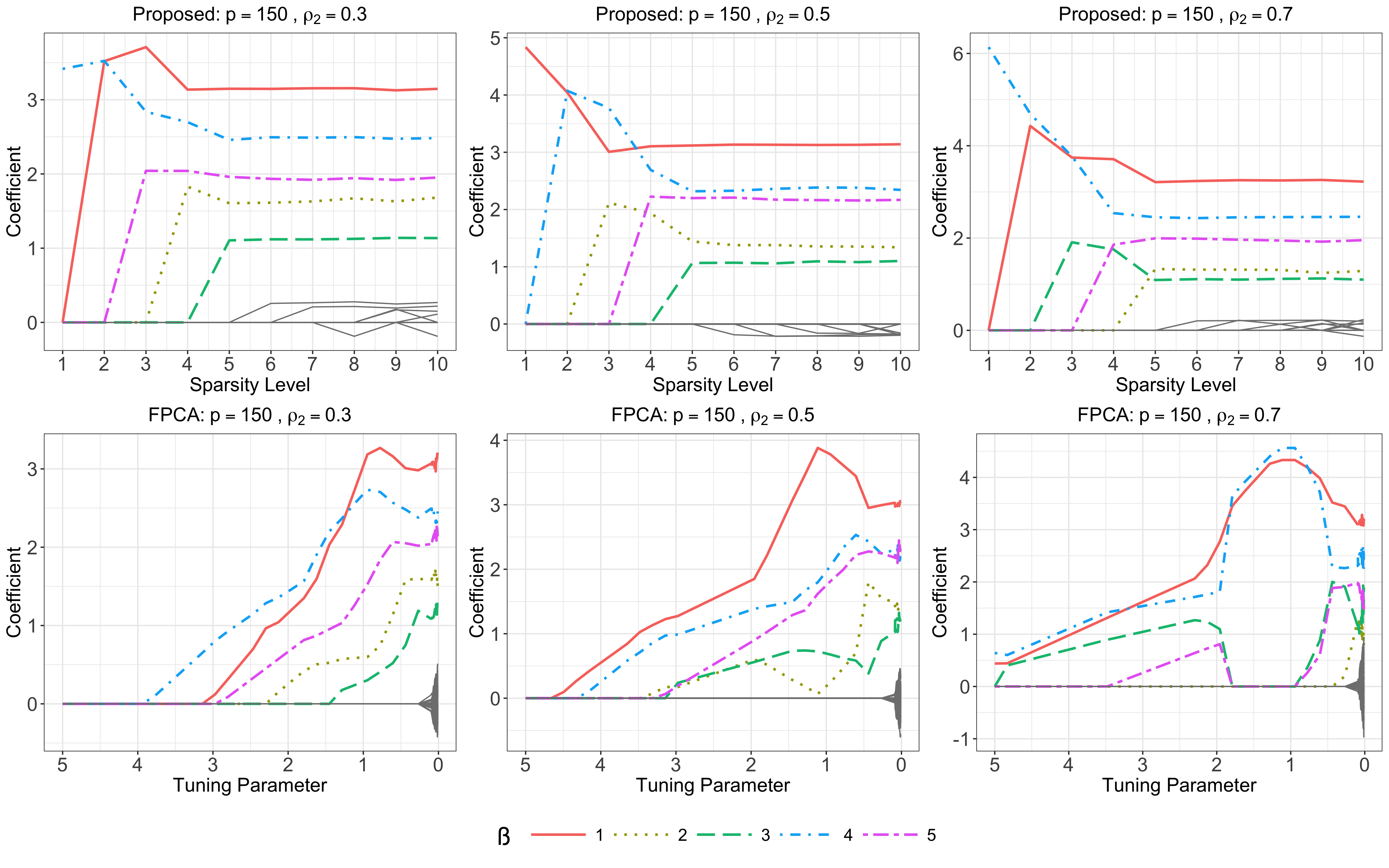}
	\caption[Solution paths of $\widehat{\beta}$]{\small Solution paths of $\widehat{\beta}$ for different settings of for different $\rho_2$ corresponding to $(p, C_0, \rho_1)=(150, 1, 0.2)$. The colored lines correspond to the solution paths of the nonzero elements of $\beta$, whereas  the grey ones correspond to zero elements. } 
	\label{fig:solpath}
\end{figure}

\end{example}

\begin{example}
In this example, we evaluate the Type I and II error rates of the proposed test statistic.
Data settings are the same as in Example \ref{ex:comparisions with yao} except that  $\xi(t) = \sum_{k=1}^{50} B (-1)^{k+1} k^{-2} \phi_k(t)$ with
$B \in \{0, 0.01, 0.03, 0.05, 0.07, 0.1\}$, which controls the signal strength.
When $B=0$, we obtain the sizes.
Because the testing results have similar patterns for different values of $(\rho_1, \rho_2)$, we report the sizes and powers when $(\rho_1, \rho_2)=(0.2, 0.5)$ for the sake of a concise presentation.
We choose $n \in \{200, 400\}$, $p \in \{150, 1500\}$ and the significance level to be 5\%.

For the null hypothesis, $H_0: \xi(t)=0$, Table \ref{table:testing_functional} summarizes the sizes and powers of the proposed test based on 1000 simulation runs. It reveals that 
the empirical sizes are reasonably controlled around the nominal level, and the empirical power increases with the sample size $n$ as well as the signal strength.

\begin{table}
	\begin{center}
		\caption{ Testing results of Monte Carlo averages with $\rho_1=0.2, \rho_2=0.5$.}
		\label{table:testing_functional}
		
		\vspace{1ex}
		\scriptsize
		\tabcolsep 4pt
		\renewcommand{\arraystretch}{1.4}
		\begin{tabular}{cccccccccccc}
			\hline
			\hline
			$n$  &      $p$    &  $B=0$ & $B=0.01$ & $B=0.03$ & $B=0.05$ & $B=0.07$ & $B=0.1$ \\
			200   & 150  & 0.057 & 0.111 & 0.112  & 0.817 & 0.969 & 1 \\
			\cline{2-8}
			& 1500 & 0.053 & 0.117 & 0.117  & 0.784 & 0.955 & 0.999\\
			\hline
			400 & 150  & 0.055 & 0.137 & 0.137 & 0.982  & 0.999 & 1 \\
			\cline{2-8}
			& 1500  & 0.057 & 0.115 & 0.115 & 0.978 & 0.999 & 1 \\
			\hline \hline
		\end{tabular}
	\end{center}
\end{table}
\end{example}

\section{ADNI Data Analysis} \label{sec:PFL-RealData}



\subsection{ GIC Pathways }
\label{sec:real_data_DC_projection}

We denote model  
\eqref{model:PFL} in this subsection as Model 1.
We treat  the  hippocampus morphometry surface  as the two-dimensional  function $Z_i(\cdot,\cdot)$.
We  also  consider the following demographic covariates. 
Specifically, $X_i$ includes
the  allele codes of screened SNPs,   the  set of demographic covariates at baseline detailed in Section \ref{sec:PFL-DataDes}, and the top 5 principal components (PCs) of the whole genome data for correcting for population stratification  \citep{price2006principal}.
Since 
the number of SNPs  is significantly larger than the sample size,   we first apply the sure independence screening approach   \citep{fan2008sure} to reduce the number of candidate SNPs, while controlling the demographic variables and the top 5 PCs.   We sort the SNPs in the decreasing order of their absolute correlations with each cognitive score and keep the top 1000 for each cognitive score. We combine the top 1000 SNPs of each score across all the 13 scores,  leading to  10546 different SNPs in $X_i$. 
The $Y_i$ is  one of the 13 cognitive scores at Month12 in Table S1, leading to 13 PFLMs.
For easy comparison, we standardize all SNPs, cognitive  scores and demographic variables, including Age and Years of education.
As both left and right hippocampi have 2D radial radial distance measures and the two parts of hippocampi have been found to be asymmetric \cite{pedraza2004asymmetry}, 
we apply our method to the left and right hippocampi separately.
In the estimation procedure, the controlling variables are not penalized and are always in the active set.
The initial value is set to be zero. We choose the sparsity level $J$ and the smoothness parameter $J$
by a grid search method with $J \in \{1, 2, \dots, 100 \}$ and $\lambda$ from 50 evenly spaced points on  $[1e-5, 0.1]$.

Figs \ref{fig:est_DC_projection_positive} and \ref{fig:est_DC_projection_negative}, respectively,  present 
the estimates of the left and right hippocampal surfaces for all the thirteen cognitive scores in Model 1. 
Fig \ref{fig:est_DC_projection_positive} shows 
the estimates  $\widehat \xi$, most of whose values range from 0.071 to 0.63, corresponding to  DIGITSCOR,  LDELTOTAL, MMSE, RAVLT.immediate,  and RAVLT.learning.
Fig \ref{fig:est_DC_projection_negative} shows 
the estimates  $\widehat \xi$, most of whose values range from -0.62 to -0.018, corresponding to  ADAS11, ADAS13, ADASQ4, CDRSB, FAQ, RAVLT.forgetting, RAVLT.perc.forgetting, and TRABSCOR.
Inspecting 
Figs \ref{fig:est_DC_projection_positive} and \ref{fig:est_DC_projection_negative} reveals the heterogeneous effects of  the hippocampus on all the 13 cognitive scores.
A bilateral and asymmetric hippocampal effect on the cognitive function is also observed.
Among the six hippocampal subfields in Fig \ref{fig:est_DC_projection_positive},  
CA1, presubiculum and subiculum show high sensitivity 
\citep{frisoni2008mapping,  de2015structural}. 
AD-related atrophy is initially focal in CA1 before spreading to other subfields.


Hereafter, we focus on  the results using the left hippocampal surface data.
Table \ref{table:est_all}  presents the  estimates of selected important covariates
and their corresponding raw $p$-fvalues for the 13 cognitive scores in Model 1.
The $p$-values are obtained using the limit distribution in Theorem 3 of the supplementary material by taking 
the true value $\beta_{A^*}^* $ to be zero. Under the null hypothesis $H_0: \beta_{A^*}^* =0$, each element of $\widehat{\beta}_{A^*}$ converges to a normal distribution and we can use the corresponding null limit distribution to calculate the $p$-value.
In consist with the existing literature \citep{vina2010women, guerreiro2015age,zhang1990prevalence},
age and  education are significant for most of the cognitive scores. 
Generally, age produces a negative effect on the cognitive function.  
Education exhibits negative effects on scores with higher value indicating impairment as well as positive effects on scores with lower value indicating impairment.
Retirement is significant for 6 scores, suggesting  an excess risk of cognition deficit among retired individuals. 
Gender is significant for 5 scores and Handedness is significant for 3 scores. 


\begin{table}
\caption{ADNI data analysis results: estimates with their corresponding raw $p$-values in parentheses of some selected covariates  for the 13 cognitive scores.}
\begin{center}
	\tiny
	\tabcolsep 2pt
	\begin{tabular}{lcccccccccccccc}
		\hline \hline
		Score  &   Model &   Gender  &  Handedness  &   Education   &  Retirement  & Age  & APOE4  & MCI   &  AD    \\ 
		\midrule	
		ADAS11	& Model 1	& 0.098(0.033)	& 0.110(0.204)	& -0.078(0.001)	& 0.137(0.021)	& 0.051(0.028)	& -	& -	& -	\\
		& Model 2	& 0.129(0.000)	& 0.145(0.012)	& -0.091(0.000)	& 0.070(0.088)	& 0.074(0.000)	& 0.252(0.000)	& -	& -	\\
		& Model 3	& 0.136(0.000)	& 0.051(0.314)	& -0.017(0.226)	& -0.019(0.585)	& 0.062(0.000)	& 0.066(0.000)	& 0.823(0.000)	& 1.855(0.000)	\\
		\hline
		ADAS13	& Model 1	& -0.034(0.333)	& 0.123(0.057)	& -0.124(0.000)	& 0.007(0.872)	& 0.050 (0.005)	& -	& -	& -	\\
		& Model 2	& 0.099(0.005)	& 0.084(0.207)	& -0.176(0.000)	& 0.158(0.000)	& 0.040(0.023)	& 0.291(0.000)	& -	& -	\\
		& Model 3	& 0.164(0.000)	& 0.004(0.929)	& 0.028(0.021)	& 0.010(0.752)	& 0.043(0.001)	& 0.077(0.000)	& 0.994(0.000)	& 1.967(0.000)	\\
		\hline
		ADASQ4	& Model 1	& -0.058(0.100)	& 0.161(0.016)	& -0.138(0.000)	& 0.042(0.348)	& -0.035(0.051)	& -	& -	& -	\\
		& Model 2	& -0.026(0.452)	& 0.135(0.037)	& -0.169(0.000)	& 0.100(0.026)	& 0.007(0.698)	& 0.308(0.000)	& -	& -	\\
		& Model 3	& 0.007(0.757)	& 0.141(0.002)	& -0.091(0.000)	& -0.001(0.969)	& 0.006(0.612)	& 0.173(0.000)	& 1.118(0.000)	& 1.746(0.000)	\\
		\hline
		CDRSB	& Model 1	& 0.019(0.660)	& 0.097(0.220)	& -0.091(0.000)	& 0.132(0.017)	& 0.047(0.027)	& -	& -	& -	\\
		& Model 2	& 0.085(0.011)	& -0.0.089(0.172)	& -0.081(0.000)	& 0.265(0.004)	& 0.053(0.002)	& 0.215(0.000)	& -	& -	\\
		& Model 3	& 0.107(0.000)	& -0.029(0.494)	& 0.016(0.154)	& 0.107(0.000)	& 0.046(0.000)	& 0.071(0.000)	& 0.819(0.000)	& 2.017(0.000)	\\
		\hline
		FAQ	& Model 1	& 0.050(0.232)	& 0.052(0.519)	& -0.079(0.000)	& 0.278(0.000)	& 0.045(0.035)	& -	& -	& -	\\
		& Model 2	& 0.145(0.004)	& 0.020(0.830)	& -0.067(0.007)	& 0.273(0.000)	& 0.002(0.925)	& 0.238(0.000)	& -	& -	\\
		& Model 3	& 0.125(0.000)	& 0.038(0.429)	& 0.019(0.135)	& 0.128(0.000)	& 0.012(0.382)	& 0.118(0.000)	& 0.666(0.000)	& 1.952(0.000)	\\
		\hline
		RAVLT.forgetting	& Model 1	& -0.026(0.580)	& 0.002(0.979)	& -0.043(0.056)	& -0.297(0.000)	& -0.158(0.000)	& -	& -	& -	\\
		& Model 2	& -0.072(0.055)	& -0.231(0.001)	& -0.033(0.075)	& -0.256(0.000)	& -0.161(0.000)	& 0.148(0.000)	& -	& -	\\
		& Model 3	& 0.062(0.219)	& -0.120(0.202)	& 0.002(0.953)	& -0.173(0.008)	& -0.047(0.064)	& 0.024(0.353)	& 0.637(0.000)	& 0.539(0.000)	\\
		\hline
		RAVLT.perc.forgetting	& Model 1	& -0.106(0.005)	& 0.120(0.081)	& -0.149(0.000)	& -0.019(0.685)	& -0.045(0.018)	& -	& -	& -	\\
		& Model 2	& -0.107(0.001)	& 0.295(0.000)	& -0.125(0.000)	& -0.009(0.830)	& -0.047(0.003)	& 0.188(0.000)	& -	& -	\\
		& Model 3	& -0.054(0.052)	& 0.022(0.659)	& -0.054(0.000)	& -0.170(0.000)	& 0.003(0.855)	& 0.093(0.000)	& 0.989(0.000)	& 1.481(0.000)	\\
		\hline
		TRABSCOR	& Model 1	& -0.036(0.436)	& -0.088(0.293)	& -0.230(0.000)	& -0.021(0.712)	& 0.003(0.876)	& -	& -	& -	\\
		& Model 2	& 0.004(0.926)	& -0.131(0.076)	& -0.182(0.000)	& 0.023(0.065)	& 0.042(0.037)	& 0.180(0.000)	& -	& -	\\
		& Model 3	& 0.010(0.803)	& -0.012(0.867)	& -0.143(0.000)	& 0.000(1.000)	& 0.061(0.002)	& 0.052(0.011)	& 0.642(0.000)	& 1.484(0.000)	\\
		\hline
		DIGITSCOR	& Model 1	& 0.189(0.000)	& 0.038(0.584)	& 0.227(0.000)	& 0.151(0.002)	& -0.076(0.000)	& -	& -	& -	\\
		& Model 2	& 0.253(0.000)	& -0.118(0.057)	& 0.192(0.000)	& -0.051(0.230)	& -0.088(0.000)	& -0.176(0.000)	& -	& -	\\
		& Model 3	& 0.238(0.000)	& -0.170(0.010)	& 0.130(0.000)	& 0.021(0.639)	& -0.052(0.004)	& -0.036(0.058)	& -0.512(0.000)	& -1.389(0.000)	\\
		\hline
		LDELTOTAL	& Model 1	& 0.077(0.067)	& -0.209(0.009)	& 0.187(0.000)	& -0.024(0.659)	& -0.004(0.846)	& -	& -	& -	\\
		& Model 2	& 0.293(0.000)	& -0.159(0.006)	& 0.199(0.000)	& 0.100(0.010)	& -0.035(0.025)	& -0.349(0.000)	& -	& -	\\
		& Model 3	& 0.021(0.396)	& -0.059(0.189)	& 0.105(0.000)	& -0.052(0.090)	& -0.034(0.007)	& -0.135(0.000)	& -1.329(0.000)	& -1.868(0.000)	\\
		\hline
		MMSE	& Model 1	& -0.043(0.215)	& 0.302(0.00)	& 0.071(0.000)	& -0.121(0.008)	& -0.049(0.006)	& -	& -	& -	\\
		& Model 2	& -0.006(0.847)	& 0.240(0.000)	& 0.141(0.000)	& -0.161(0.000)	& -0.008(0.609)	& -0.230(0.000)	& -	& -	\\
		& Model 3	& -0.159(0.000)	& 0.146(0.002)	& 0.019(0.130)	& -0.080(0.011)	& -0.038(0.003)	& -0.103(0.000)	& -0.670(0.000)	& -1.752(0.000)	\\
		\hline
		RAVLT.learning	& Model 1	& 0.283(0.000)	& -0.102(0.222)	& 0.123(0.000)	& 0.048(0.401)	& -0.031(0.158)	& -	& -	& -	\\
		& Model 2	& 0.268(0.000)	& 0.092(0.018)	& 0.124(0.000)	& -0.002(0.972)	& -0.047(0.010)	& -0.227(0.000)	& -	& -	\\
		& Model 3	& 0.139(0.000)	& 0.122(0.047)	& 0.079(0.000)	& -0.059(0.160)	& 0.018(0.296)	& -0.045(0.009)	& -0.910(0.000)	& -1.240(0.000)	\\
		\hline
		RAVLT.immediate	& Model 1	& 0.425(0.000)	& 0.169(0.050)	& 0.228(0.000)	& 0.022(0.708)	& -0.049(0.033)	& -	& -	& -	\\
		& Model 2	& 0.457(0.000)	& 0.094(0.173)	& 0.227(0.000)	& 0.001 (0.980)	& -0.039(0.034)	& -0.268(0.000)	& -	& -	\\
		& Model 3	& 0.287(0.000)	& -0.008(0.855)	& 0.147(0.000)	& -0.134(0.000)	& -0.008(0.531)	& -0.106(0.000)	& -0.973(0.000)	& -1.602(0.000)	\\ 	
		\hline \hline
	\end{tabular}
	\label{table:est_all}
\end{center}
\tiny{ Model 1 corrects  for all covariates except APOE4 and the baseline disease status; Model 2 corrects for all covariates except the baseline disease status; and Model 3 corrects for all covariates.
}
\end{table}


\begin{figure}
\centering
\centering
\includegraphics[width=1 \linewidth]{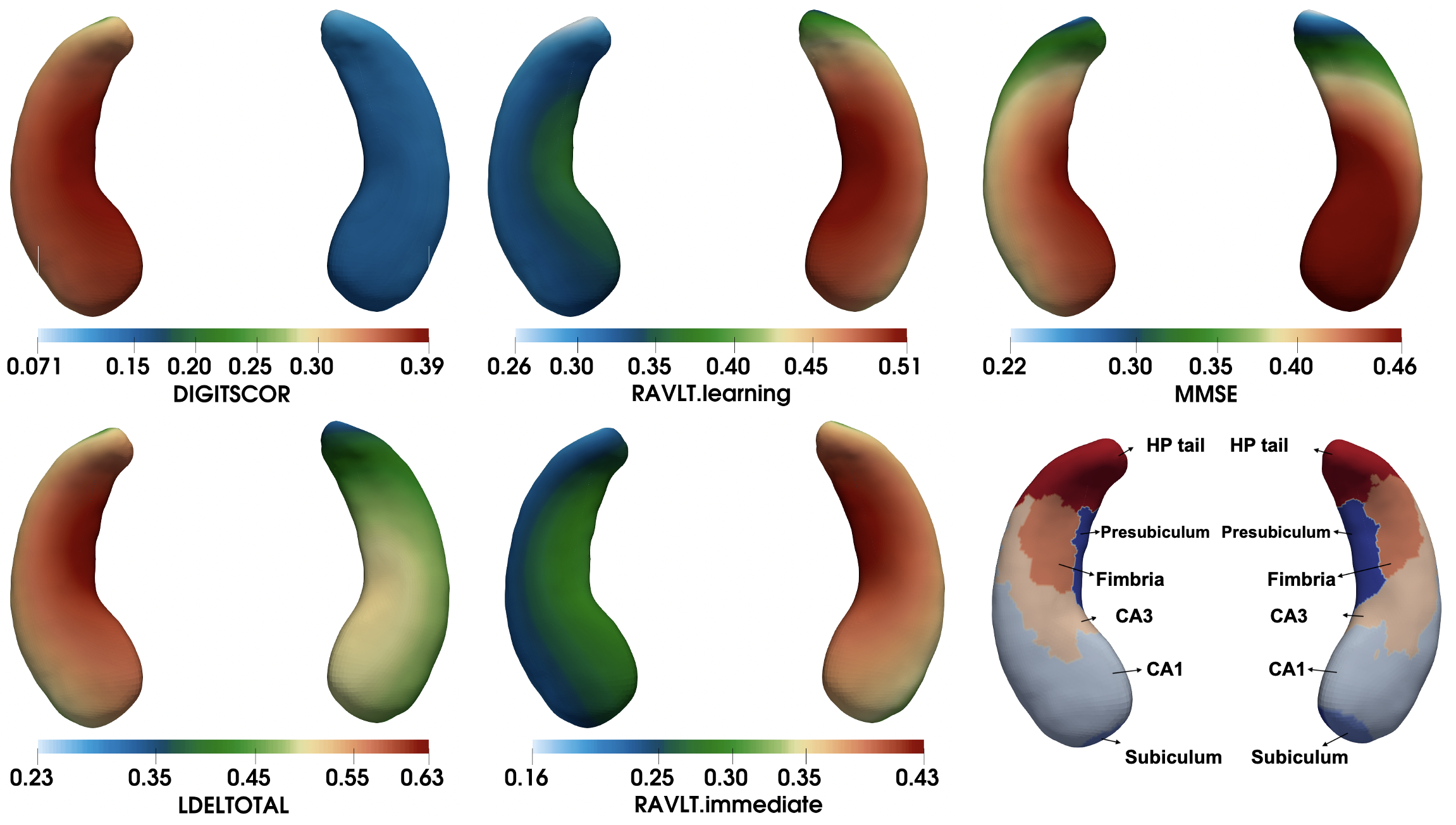}
\caption{ADNI data analysis results:  estimates of the left and right hippocampus surfaces for   DIGITSCOR,  LDELTOTAL, MMSE,
	RAVLT.immediate and RAVLT.learning
	for Model 1 and the hippocampal subfields (from left to right, and from top to bottom).}
\label{fig:est_DC_projection_positive}
\end{figure}

\begin{figure}
\centering
\includegraphics[width=1 \linewidth]{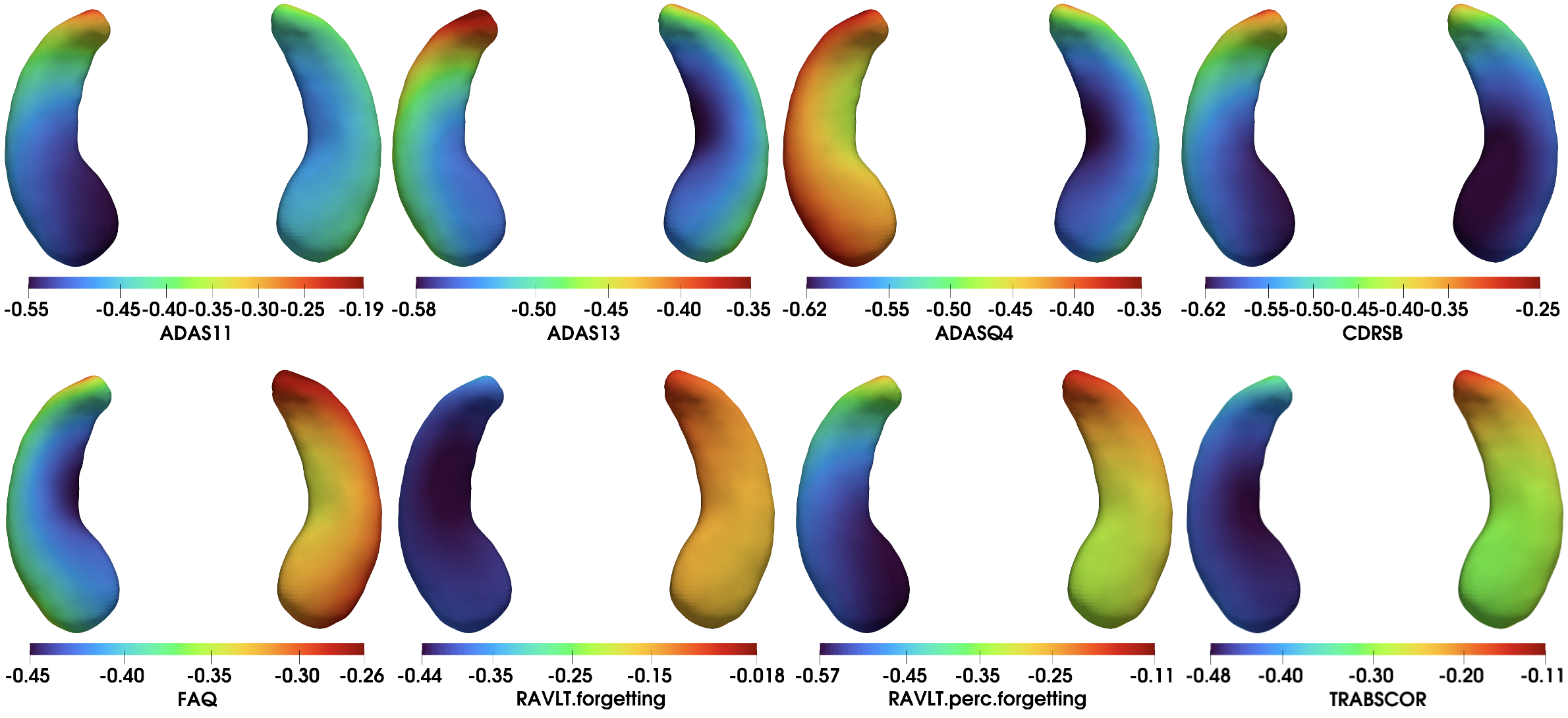}
\caption{ADNI data analysis results:   estimates of the the left and right hippocampus surfaces for  ADAS11, ADAS13, ADASQ4, CDRSB, FAQ,
	RAVLT.forgetting, RAVLT.perc.forgetting and TRABSCOR
	for Model 1 (from left to right, and from top to bottom).}
\label{fig:est_DC_projection_negative}
\end{figure}

Fig \ref{fig:pheno_gram_DC_projection} and  Fig \ref{fig:Multraits_pvalues_DC} present  
the ideogram  and   Manhattan plots of  significant SNPs for all the 13 scores in  Model 1, respectively. 
Inspecting Fig \ref{fig:pheno_gram_DC_projection} reveals heterogeneous genetic effects across
all the scores, but 
several well-known SNPs on the 19th chromosome are identified to be important for all  the 13 cognitive scores.
Table \ref{table:snp} lists  SNPs on the 19th chromosome identified to be important for at least 3 cognitive scores. 
The rs429358 on the 19th chromosome, which is one of the two variants for the well-known APOE alleles,  is significant for   
ADAS11, ADAS13, FAQ, MMSE, RAVLT.immediate, RAVLT.perc.forgetting.
Other SNPs include  
rs283812, in the PVRL2 region,
rs769449 in the APOE region,
rs66626994 in the APOC1 region. 
The  APOE, PVRL2 and APOC1 regions in the cytogenetic region 19q13.32 are high AD-risk regions
\citep{carrasquillo2009genetic, vermunt2019duration}.

Except for the 19th chromosome,  two SNPs are also identified to be important for at least 3 scores, including
rs28414114 from chromosome 14 with the smallest $p$-value 7e-14 and
rs12108758 from chromosome 5 with the smallest $p$-value 2e-9.
Furthermore, we obtain the $p$-values of the selected SNPs in Fig \ref{fig:Multraits_pvalues_DC}.
The $p$-values of several important SNP are smaller than $0.05/10145$ (10145 SNPs after screening).




\begin{figure}
\centering
\subfigure[]{\label{fig:pheno_gram_DC_projection}\includegraphics[width=0.8\linewidth]{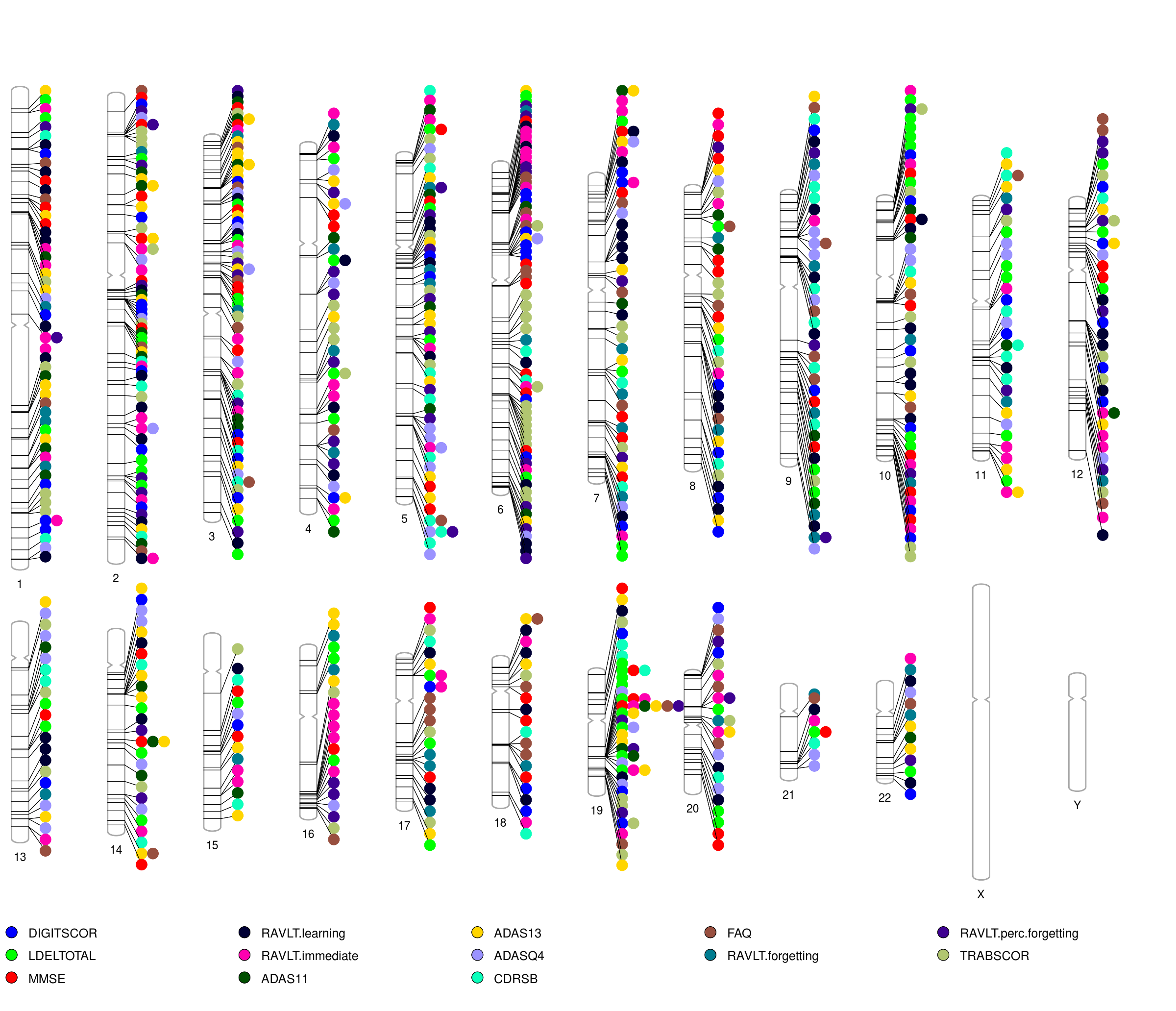} } \\
\subfigure[]{\label{fig:Multraits_pvalues_DC}\includegraphics[width=0.9\linewidth]{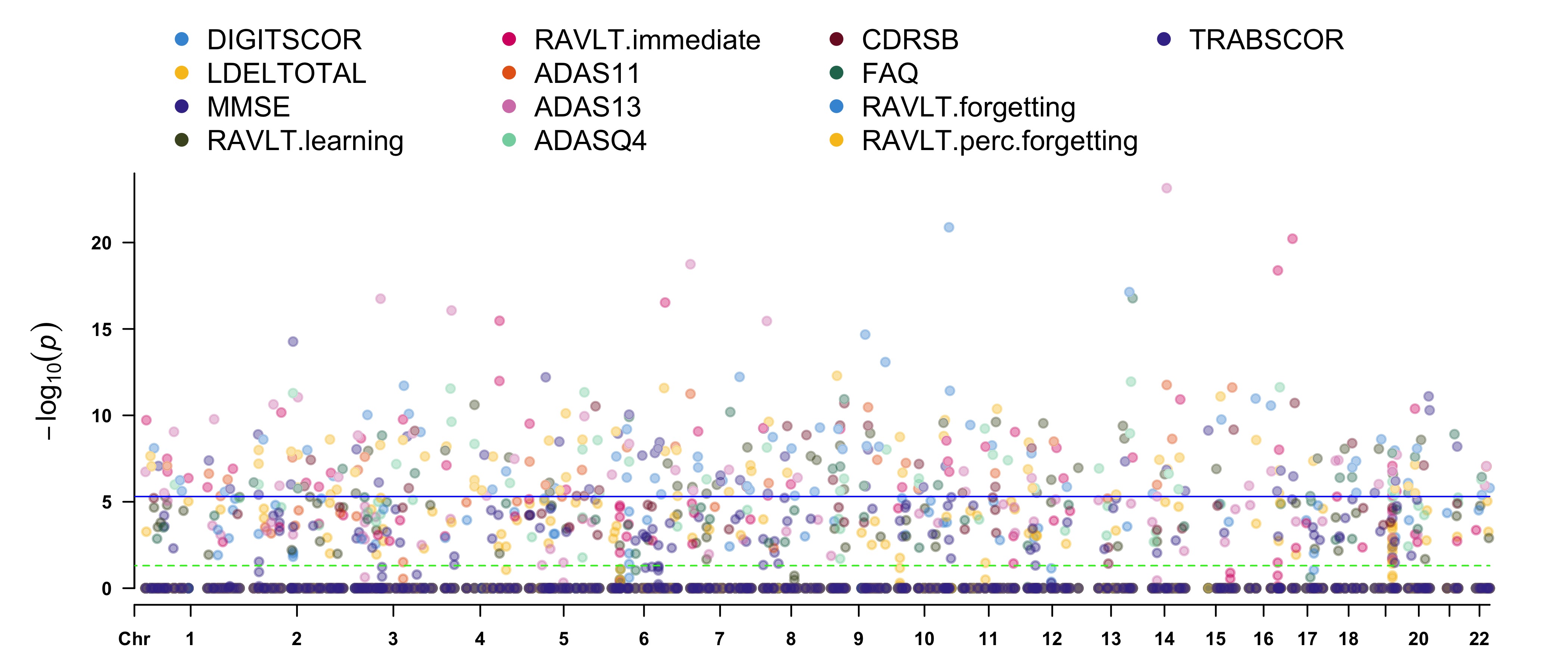} } 
\caption{ Panel (a) presents  positions of the important SNPs for the 13 scores for Models 1. The colors represent the 13 scores and each signal dot indicates the SNP is selected for the corresponding score. Panel 
	(b) gives $p$-values of the selected SNPs for Model 1. The blue line indicates the threshold $p$-value$=0.05/10154$.}
\label{fig:pheno_grams_all_models}
\end{figure}


\begin{table}
\caption{Detailed information of the common SNPs from the 19th chromosome
	for at least 3 scores for Model 1.}
\centering
\scriptsize
\tabcolsep 4pt
\renewcommand{\arraystretch}{1.4}
\begin{tabular}{lrrl}
	\hline \hline
	\multicolumn{1}{c}{SNP}&\multicolumn{1}{c}{Chr}&\multicolumn{1}{c}{Base Pair}  & Scores  \tabularnewline
	\hline
	rs283812   & 19  &  45388568    & CDRSB, LDELTOTAL, MMSE  \\  
	rs769449   &  19   &  45410002 &  LDELTOTAL, MMSE, RAVLT.immediate  \\ 
	rs429358   &  19  & 45411941    & ADAS11, ADAS13, FAQ, MMSE, RAVLT.immediate, RAVLT.perc.forgetting \\ 
	rs66626994  &  19   & 45428234  &  ADAS13, LDELTOTAL, RAVLT.immediate\\  
	\hline \hline
\end{tabular}
\label{table:snp}
\end{table}

\subsection{Conditional GIC (CGIC) pathways given APOE4}
\label{sec:real_data_DC_apoe}

We denote model  
\eqref{model:PFL} in this subsection as Model 2.
Model 2 is  almost the same as Model 1 except that 
we  exclude the SNPs in the 19q13.32 region from the candidate SNPs and include the number of APOE4 gene copies as one of the controlling covariates. 
In our dataset,   230 subjects had one APOE4 allele and 67 subjects had two APOE4 alleles. 
The cytogenetic region 19q13.32 contains 6376 SNPs in this region, including  
the well-known APOE  \citep{bertram2012genetics, zhao2021genetic}. It allows us to  better understand the conditional effects of other SNPs on cognitive scores given  on the APOE4 alleles. 
We apply the same screening step and Algorithm \ref{alg:estimation} to Model 2 for each cognitive score.

Table \ref{table:est_all}  presents the related estimation results corresponding to    Model 2.
Estimates of the demographic covariates in Model 2 are similar to their corresponding estimates in Model 1.
The number of APOE4 alleles is significant for all the scores, while exhibiting  negative effects on cognitive ability. 
Estimates of the hippocampal surface for the 13 scores in Model 2 are similar to those 
in Model 1, so we include them  in the supplementary material.

Fig S5(a) of the supplementary material presents the ideogram of the selected important SNPs for Model 2.
For each cognitive score, the selected significant SNPs  in Model 2 enjoy some similarities with those in Model 1. For instance,  
rs28414114 from chromosome 14 are also identified to be important for ADAS11 and ADAS13.
Meanwhile, regions of the selected SNPs for some scores are similar to that in Model 1.
For example, positions of the important SNPs for TABSCOR in chromosome 6 are from 81858848 to 91078328 with the smallest $p$-value 9.06e-11
and from 120546293 to 120765041 with
the smallest $p$-value 8e-9 in Model 1,
and in Model 2 the important positions are from 80949537 to 94053838 with
the smallest $p$-value 1.16e-9
and from 120533534 to 120771669 with
the smallest $p$-value 6e-15.


\subsection{CGIC  pathways given APOE4 and disease status}

We denote model  
\eqref{model:PFL} in this subsection as Model 3.
Model 3 is  almost the same as Model 2 except that 
we further  include the baseline diagnosis status as one of the controlling covariates. 
The baseline diagnosis status is coded by using two  dummy variables: MCI and AD. 
Because clinical notes provide supplementary information and are considered on a case-by-case basis,
the effects of the SNPs on change in cognitive performance may be confounded with the effects 
of differences in baseline diagnosis.
We are interested in whether the relationships would alter when adjusting for the baseline
diagnosis status.
We apply the same screening step and Algorithm \ref{alg:estimation} to Model 3 for each cognitive score.

Table \ref{table:est_all} also  presents the related estimation results corresponding to Model 3. 
After introducing the baseline diagnosis status,
almost all estimates of 
the demographic covariates and APOE4 in Model 3 are smaller than their corresponding estimates in Models 1 and 2. 
The baseline status MCI has significant positive effects on ADAS11, ADAS13, CDRSB, FAQ, RAVLT.forgetting, RAVLT.perc.forgetting and TRABSCOR,
and exhibit significant negative effects on DIGITSCOR,  LDELTOTAL, MMSE, RAVLT.learning and RAVLT.immediate.
The baseline status AD generally has stronger effects on the 13 scores in Month 12 than the baseline MCI status.
Similar patterns of the hippocampal estimates are also observed
for the 13 scores to those in Section \ref{sec:real_data_DC_projection} and we include the corresponding results in the supplementary material.

Fig S5(b) of the supplementary material presents the ideogram of the selected important SNPs for Model 3.
The selected important SNPs seem quite different to the SNPs in Fig \ref{fig:pheno_gram_DC_projection}.
This is reasonable because we consider the baseline diagnosis status in the screening step and always keep it in the model.
The selected important SNPs for at least 3 scores are 
rs13101604 from chromosome 4 with the smallest $p$-value 4.18e-9,
rs2442696 from chromosome 4 with the smallest $p$-value 6.96e-11, 
and rs4761161 from chromosome 12 with the smallest $p$-value 2.11e-09.

\subsection{Comparisons of the three models}
In this subsection, we compare the above three models in terms of the shared and different
heritability patterns of the 13 scores and the proportions of the variations explained
in cognitive deficits by the three types of data:
the genetic data, the controlling covariates and the hippocampal surface data.
The average computation times of the proposed method are 2.25 hours with the standard error 0.41 hour
for Model 1, 2.31 hours with the standard error 0.44 hour for Model 2, and 2.19 hours with the standard error 0.45 hour for Model 3.

Although most human traits have a polygenic architecture \citep{wray2018common}, heritability 
can be used to measure how much of the variation in each score is due to variation in genetic data \citep{yang2010common}.
By definition, we estimate the heritability for the three models by calculating the phenotypic variance due to the genetic variables.
We estimate the phenotypic variance by the empirical variance of $X_{iG}^\top \hat \beta_G $, where $X_{iG}$ is the genetic variables of the $i$-th subject and $\hat \beta_G$ is the corresponding coefficient estimates.
For Model 2 and Model 3, estimates of the heritability are based on the considered SNPs which excludes the SNPs in 
the cytogenetic region 19q13.32.
Fig \ref{fig:Heritability_estimates} gives the heritability estimates of the genetic effects
for the cognitive scores.
Heritability of the cognitive scores are estimated to be 62.69\%$\sim$78.01\% for Model 1, and
55.56\%$\sim$85.62\% for Model 2, and 33.02\%$\sim$69.66\% for Model 3.
The remaining heritability of the scores, especially RAVLT.forgetting, is still relatively high
even after accounting for APOE4. It is consistent with previous research 
and suggests that memory functioning in AD is under strong genetic influence that is only partly attributable to APOE genotype \citep{wilson2011heritability}. 
However, there are 1.1\%$\sim$35.3\% decreases of the heritability estimates of the 13 scores for Model 3.
It reveals that the baseline diagnosis status explains a part of the cognitive function associated to the polygenic effect.

We also examine the effect size of the controlling covariates by
calculating the proportion of variance explained by these covariates in 
Fig \ref{fig:variance_clinical}. 
The proportions of variance explained by the controlling variables increase with the inclusion of the number of APOE4 alleles and the baseline disease status.

Fig \ref{fig:variance_imaging} presents the effect size of the imaging covariates.
The hippocampal surface data account for 1\% $\sim$4.6\% of the total variantions in 13 cognitive scores for Model 1, 0.1\%$\sim$4.1\% for Model 2 and 0.005\%$\sim$0.63\% for Model 3.
These results suggest that the baseline diagnosis status explains a larger part of the cognitive function associated to the hippocampal data compared to the number of APOE4 gene alleles.
We also compare the effects of the genetic data with that of the imaging data.
The above results show that the cognitive function may have the polygeneic inheritance, which
is not controlled by one gene, but by multiple genes that each make a small contribution to the overall outcome. There are 75 genes on average selected to be important for the congnitive scores in the three models. The variance explained by a single gene is about
0.84\%$\sim$1.05\%,  0.74\% $\sim$ 1.14\% and 0.44\% $\sim$ 0.93\% in the three models, which is comparable to that of the imaging data.
Furthermore, we calculate the estimated $p$-values of left and right hippocampal
surface data for the 13 cognitive scores in the three models in Table \ref{table:image_pvalue}.
The left hippocampus is significant at 5\% significance level for 9 scores, 11 scores and
5 scores in Model1, Model 2 and Model 3, respectively.
The right one is significant for 7 scores,  11 scores, and 3 scores in Model1, Model 2 and Model 3, respectively. 
Both the left and right hippocampa are significant for ADAS13 and MMSE, which are consistent with the findings in \cite{morrison2022regional} and \cite{peng2015correlation}.
It can also be observed that most of the $p$-values in Model 3 are larger than the corresponding $p$-values in Model 1 and Model 2.


\begin{table}
\caption{Estimated $p$-values of hippocampal surface data for the 13 cignitive scores.}
\begin{center}
	\tiny
	\tabcolsep 3pt
	\renewcommand{\arraystretch}{1.5}
	\begin{tabular}{lrrrrrrrrrccccc}
		\hline \hline
		& \multicolumn{2}{c}{Model 1} & &\multicolumn{2}{c}{Model 2} & & \multicolumn{2}{c}{Model 3} \\
		\cline{2-3} \cline{5-6} \cline{8-9}
		Score  &   Left & Right &  &   Left & Right  &  &   Left & Right  \\ 
		\midrule	
		ADAS11	&	3.66E-05	&	1.35E-08 &	&	6.72E-05	&	1.04E-05&	&	0.054	&	0.046	\\
		\hline
		ADAS13	&	1.58E-08	&	1.49E-09&	&	6.36E-08	&	2.11E-05&	&	0.016	&	0.023	\\
		\hline
		ADASQ4	&	2.55E-07	&	2.82E-05&	&	3.41E-07	&	2.88E-06&	&	0.052	&	0.096	\\
		\hline
		CDRSB	&	4.56E-07	&	4.27E-06&	&	2.25E-05	&	0.008&	&	0.105	&	0.142	\\
		\hline
		FAQ	&	3.72E-05	&	0.753&	&	0.998	&	0.003&	&	0.039	&	0.102	\\
		\hline
		RAVLT.forgetting	&	0.707	&	0.999&	&	0.160	&	0.257&	&	0.999	&	0.120	\\
		\hline
		RAVLT.perc.forgetting	&	0.001	&	0.824& &	9.89E-05	&	2E-04&	&	0.132	&	0.071	\\
		\hline
		TRABSCOR	&	0.916&	0.999&	&	0.007&	0.001 &	&	0.011	&	0.090	\\
		\hline
		DIGITSCOR	&	0.036&	0.999&	&	0.001	&	0.381&	&	0.117	&	0.582	\\
		\hline
		LDELTOTAL	&	2.19E-07	&	1.82E-09&	&	0.001	&	7.22E-06&	&	0.032	&	0.160\\
		\hline
		MMSE	&	2.66E-10	&	5.35E-11&	&	0.002	&	1.06E-05&	&	0.022	&	0.039	\\
		\hline
		RAVLT.learning	&	0.998	&	7.80E-05&	&	2E-04	&	5.70E-05&	&	0.999	&	0.080	\\
		\hline
		RAVLT.immediate	&	0.999	&	0.999&	&	7.83E-05	&	4E-04&	&	0.161	&	0.329	\\
		\hline \hline
	\end{tabular}
	\label{table:image_pvalue}
\end{center}
\tiny{ Model 1 corrects  for all covariates except APOE4 and the baseline disease status; Model 2 corrects for all covariates except the baseline disease status; and Model 3 corrects for all covariates.
}
\end{table}

Genetic correlation has been proposed to describe the shared genetic associations within pairs of quantitative traits,
and it can be calculated as the correlation of the genetic effects of numerous SNPs on the traits \citep{zhao2022genetic}.
To provide critical information about the fundamental biological pathways and
describe the shared genetic etiology of the 13 scores,
we calculate the genetic correlations between the 13 scores for the 3 models that we have considered in Figs \ref{fig:Genetic_correaltion_model1}-\ref{fig:Genetic_correaltion_model3}.
There exist strong genetic correlations between the 13 scores for Model 1, which suggest the overall similarity of the genetic architecture on brain functions in Month 12
characterizing by these scores.
The genetic correlations adjusting for the number of APOE4 alleles are similar to that for Model 1 with slightly smaller values,
indicating the shared genetic effects for the 13 scores besides the effect of the well known APOE4 gene.
However, the genetic correlations decrease a lot when additionally controlling for the baseline diagnosis status, which is supposed to explain
a large part of the shared genetic effect on the cognitive scores.
Also, it reveals that greater genetic heterogeneity exists after accounting for the baseline diagnosis status. One potential reason is that the population of AD is genetically heterogeneous \citep{lo2019identification}.
Adjusting for disease status suggests that within a given diagnostic group, there is substantial variation in cognitive scores that is explained by genetic markers, suggesting heterogeneity within disease classifications that is partially explained by genetics. 
It may indicate that different therapies are needed for different symptoms after the AD onset.

\begin{figure}
\centering
\subfigure[]{\label{fig:Heritability_estimates} \includegraphics[width=0.3\linewidth]{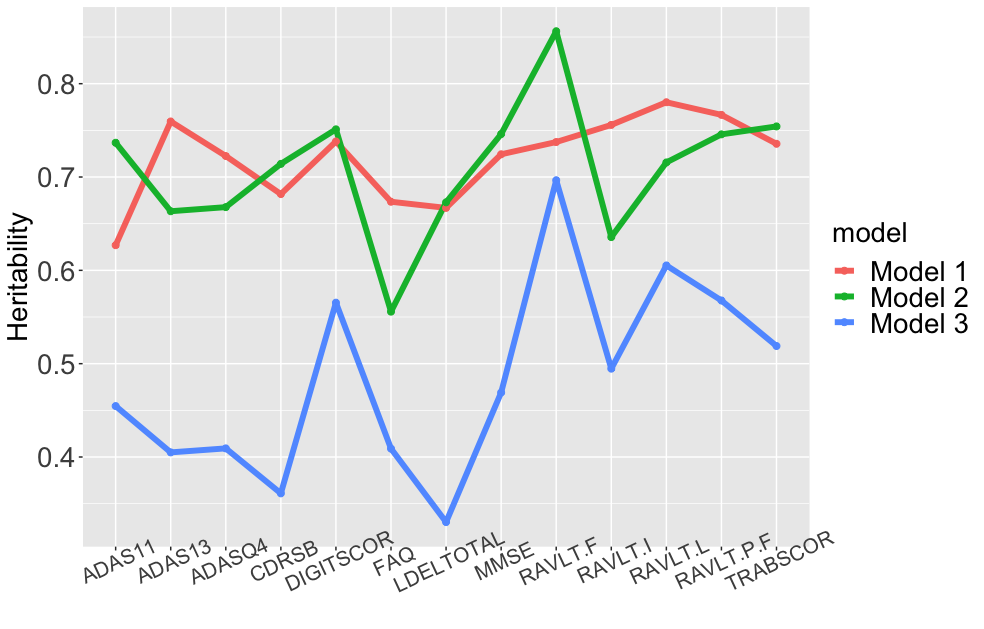}} 
\subfigure[]{\label{fig:variance_clinical} \includegraphics[width=0.3\linewidth]{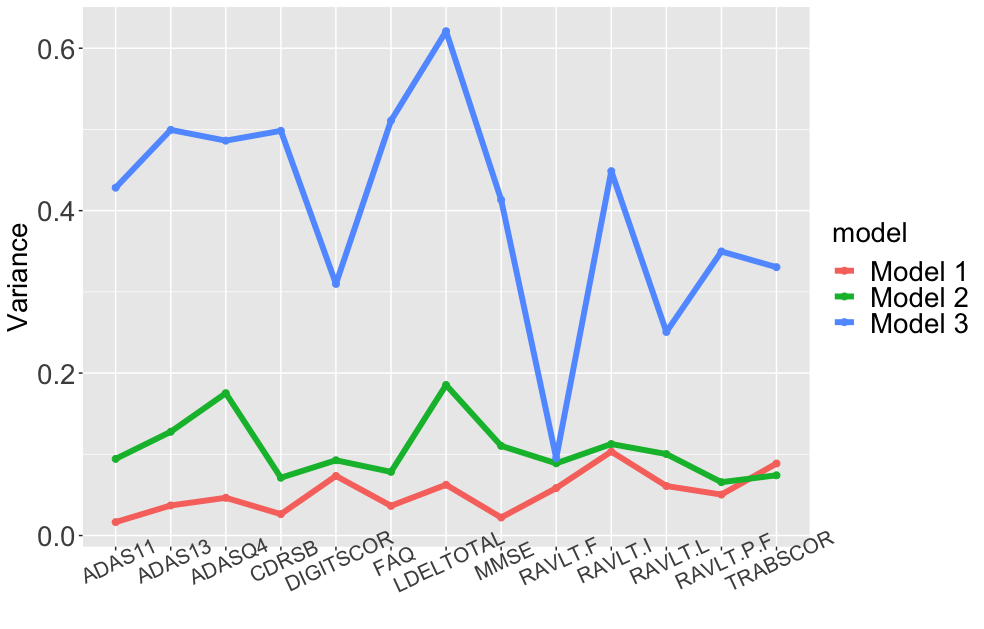} }
\subfigure[]{\label{fig:variance_imaging} \includegraphics[width=0.3\linewidth]{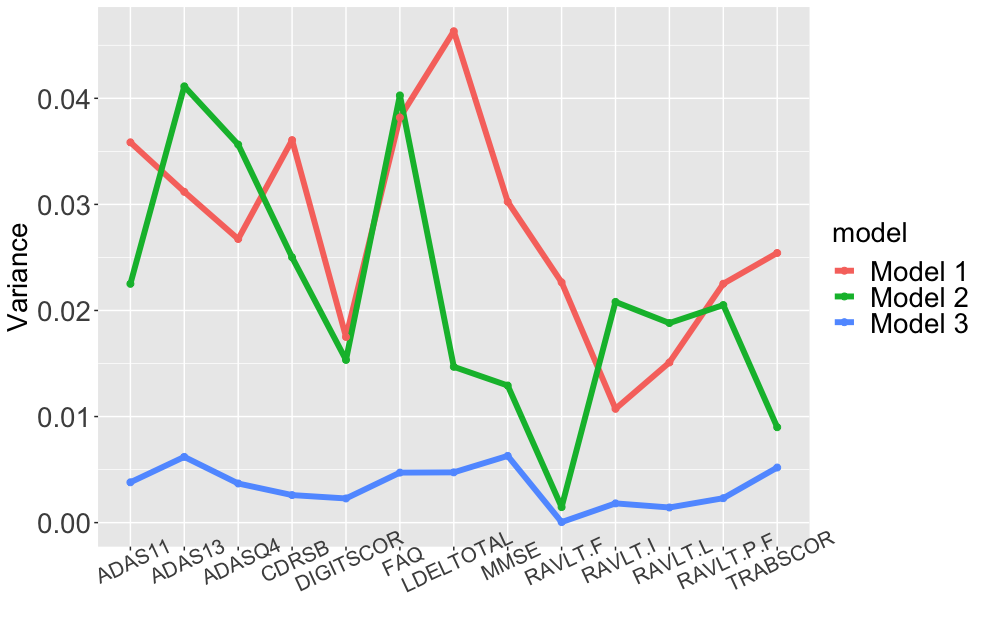} } \\
\subfigure[]{\label{fig:Genetic_correaltion_model1} \includegraphics[width=0.3\linewidth]{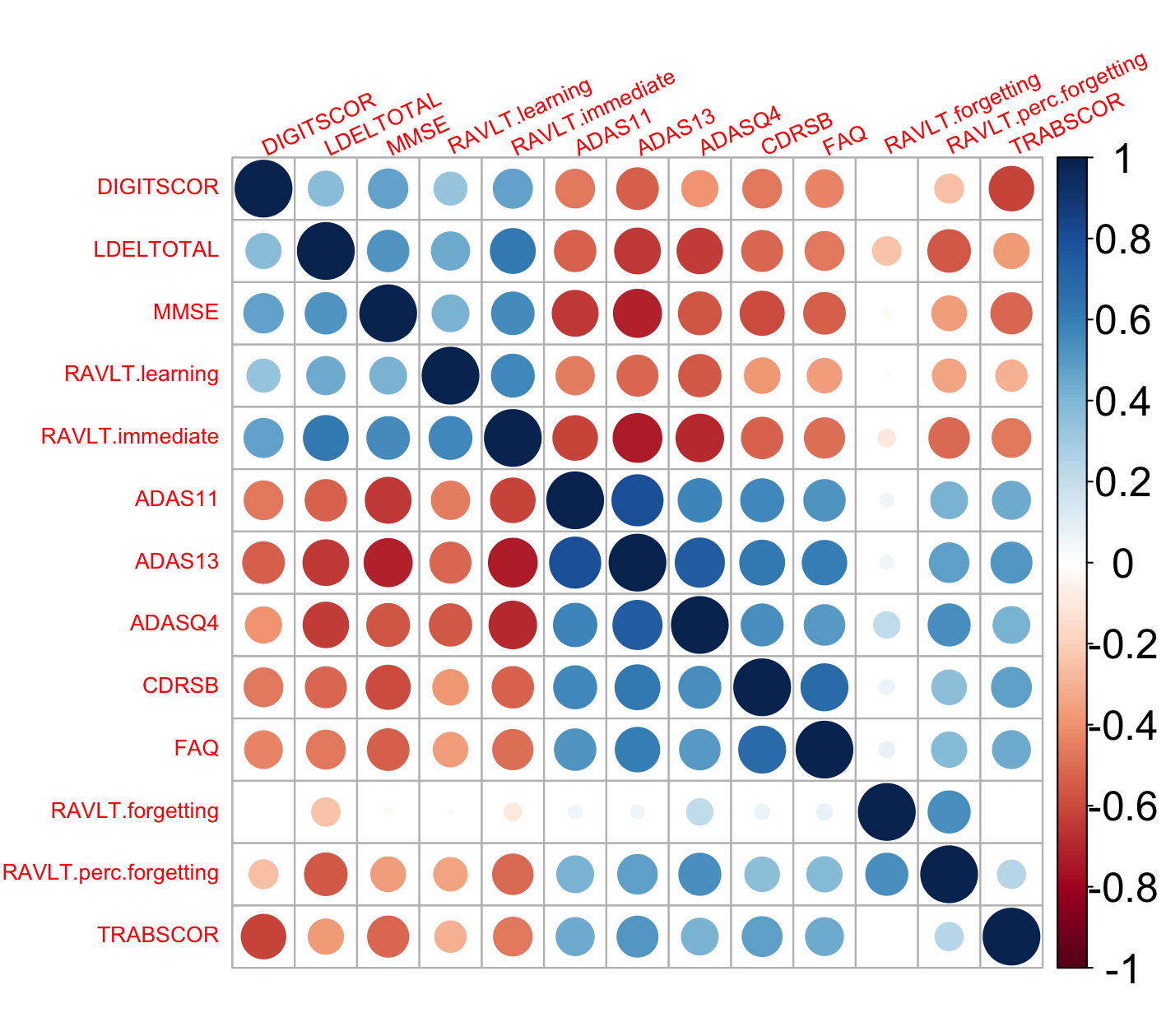} }
\subfigure[]{\label{fig:Genetic_correaltion_model2} \includegraphics[width=0.3\linewidth]{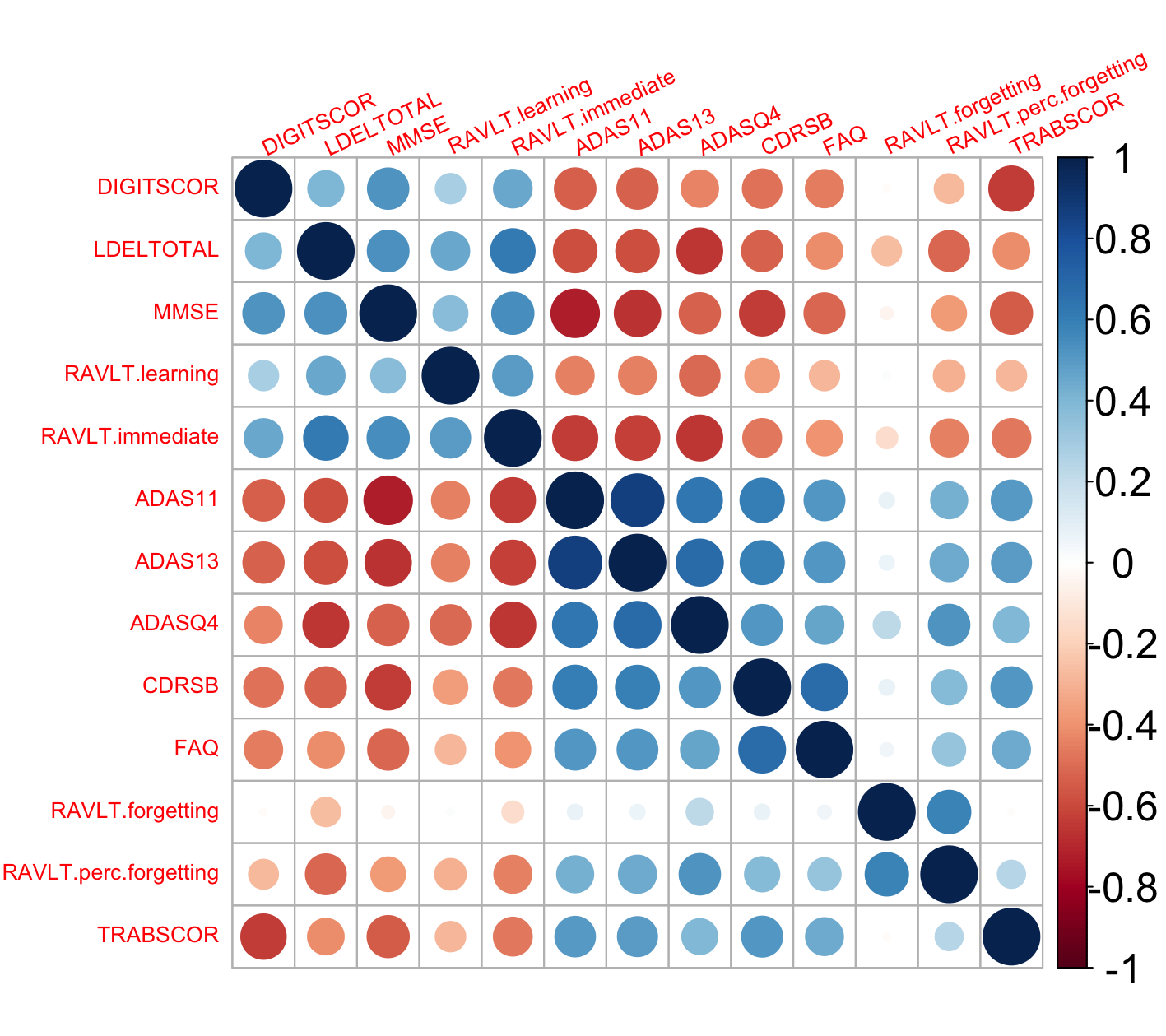} }
\subfigure[]{\label{fig:Genetic_correaltion_model3} \includegraphics[width=0.3\linewidth]{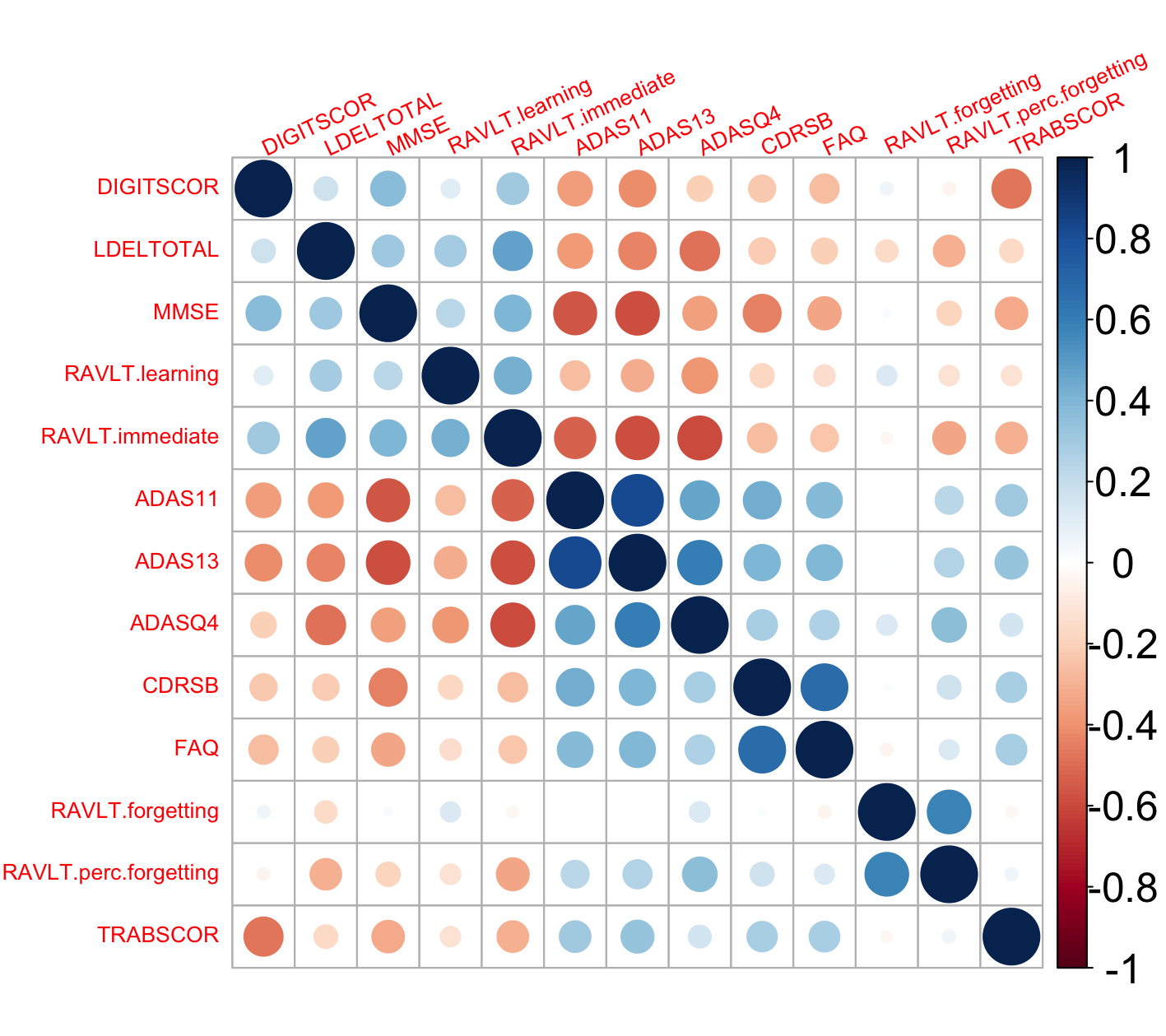}} 
\caption{
	Heritability estimates of the genetic data (a) and the proportions of variance explained by the controlling variables (b) and the imaging data (c).
	Genetic correlations between the 13 cognitive scores for Model 1 (d), Model 2 (e) and Model 3 (f). 
	The red line indicates Model 1, the green line indicates Model 2 and the blue line indicates Model 3.}
\label{fig:genetic_corr_scores}
\end{figure}



\section{Discussion}
This paper aims at mapping the biological pathways of phenotypes of interest from the ADNI study by
integrating GIC data. The high dimensional genetic data and the features of the hippocampal surface data motivate us to consider a high-dimensional PFLM to establish the associations
between the genetic and imaging data with the phenotype of interest. We proposed a new estimation method 
to high-dimensional PFLMs under the RKHS framework and the $\ell_0$ penalty, and investigated the theoretical results of the estimators. Through the analyses of the ADNI study, we have shown that 
the proposed method is a valuable statistical tool for quantifying the complex relationships between 
phenotypes of interest and the GIC data.

Although the proposed method considers one functional covariate, it can be extended to accommodate multiple functional covariates. We can consider the following high-dimensional PFLM with multiple functional predictors,
$$
Y_i = \alpha + X_i^\top \beta + \sum_{k=1}^K \int_\mathcal{T} Z_{ik} (t) \xi_k(t) dt + \epsilon_i.
$$
Each $\xi_k(t)$ is assumed to be in a reproducing kernel Hilbert space. 
Similar to the minimization problem in \eqref{eq:min} that imposes $\ell_0$ penalty to $\beta$ and RKHS roughness penalty on $\xi(t)$, we can estimate $\{ \xi_k(t)\}$s by imposing 
RKHS roughness penalty on each $\xi_k(t)$. The representer theorem and the theoretical results can be obtained similarly. Such extensions are worthy of further investigation.

\bibliographystyle{chicago}
\bibliography{RefPFLR-RKHS}

\end{document}